\documentclass[11pt,preprint]{aastex}

\usepackage{amsmath}
\usepackage{graphicx}
\usepackage{multirow}
\usepackage{color}
\usepackage{natbib}
\usepackage{url}

\def\AU{{\rm AU}}
\font\sm=cmss10 at 10pt
\graphicspath{{Figures/}}
\begin{document} 

\title{On the Non-uniform Distribution of the Angular Elements of Near-Earth Objects}

\author{Youngmin JeongAhn and Renu Malhotra}
\affil{Lunar and Planetary Laboratory, The University of Arizona, Tucson, AZ 85721, USA.}
\email{jeongahn@lpl.arizona.edu,renu@lpl.arizona.edu}

\begin{abstract}

We examine the angular distributions of near-Earth objects (NEOs) which are often regarded as uniform.   The apparent distribution of the longitude of ascending node, $\Omega$, is strongly affected by well-known seasonal effects in the discovery rate of NEOs.  The deviation from the expected $\pi$-periodicity in the apparent distribution of $\Omega$ indicates that its intrinsic distribution is slightly enhanced along a mean direction, $\bar\Omega=111^\circ$; approximately $53\%$ of NEOs have $\Omega$ values within $\pm90^\circ$ of $\bar\Omega$. We also find that each sub-group of NEOs (Amors, Apollos and Atens) has different observational selection effects which cause different non-uniformities in the apparent distributions of their arguments of perihelion $\omega$, and longitudes of perihelion $\varpi$.  For their intrinsic distributions, our analysis reveals that the Apollo asteroids have   non-uniform $\omega$ due to secular dynamics associated with inclination--eccentricity--$\omega$ coupling, and the Amors' $\varpi$ distribution is peaked towards the secularly forced eccentricity vector.  The Apollos' $\omega$ distribution is axial, favoring values near $0$ and $180^\circ$; the two quadrants centered at $0$ and $180^\circ$ account for $55\%$ of the Apollos' $\omega$ values.   The Amors' $\varpi$ distribution peaks near $\bar\varpi=4^\circ$; sixty-one percent of Amors have $\varpi$ within $\pm90^\circ$ of this peak. We show that these modest but statistically significant deviations from uniform random distributions of angular elements are owed to planetary perturbations, primarily  Jupiter's.  It is remarkable that this strongly chaotic population of minor planets reveals the presence of Jupiter in its angular distributions.

\end{abstract}

\section{Introduction}\label{s:intro}

Our knowledge of the contemporary population of near-Earth objects (NEOs) is based on many serendipitous discoveries by individuals (amateurs and professionals) as well as several dedicated sky surveys (LINEAR, NEAT, Spacewatch, LONEOS, Catalina Sky Survey(CSS), WISE, and Pan-STARRS). More than 8,000 NEOs currently have been discovered and confirmed.  
The orbits of the NEOs are known to be strongly chaotic, with frequent perturbations owed to planetary close approaches.  The typical Lyapunov times are $\sim$~100 yr~\citep{Tancredi:1998b}, whereas the dynamical lifetimes are $\sim10^7$~yr~\citep{Gladman:1997,Ito:2006}.   The NEOs also have fast (and chaotic) nodal and apsidal precession over their dynamical lifetimes.  Consequently, it may be generally expected that the longitudes of ascending node, $\Omega$, the arguments of perihelion, $\omega$, and the longitude of perihelion, $\varpi = \Omega + \omega$, of NEOs should each be randomly and uniformly distributed in the range 0 to $2\pi$.  This assumption has often been made in quantitative models of the NEOs which usually attempt to determine the distribution function, $R(a,e,i)$, for the NEOs' semimajor axis, eccentricity and inclination, while marginalizing the angular elements (e.g., \cite{Bottke:2002}).  One study where this assumption was not made was regarding the longitudes of perihelia of the Taurid asteroids \citep{Valsecchi:1999}.

 In the present paper, we examine the angular elements' distributions by use of up-to-date observational data on the NEOs. In the observed data, there are distinct non-uniformities in the $\Omega$, $\omega$, and $\varpi$ distributions, some of which have previously gone unnoticed. The distribution of $\Omega$ shares similar trends among different dynamical subgroups of NEOs, and some of these trends are related to well-known seasonal effects \citep{Jedicke:2002}.  We will show here that there is also a discernible dynamical effect in the non-uniform distribution of $\Omega$ that is owed to secular planetary perturbations. We also show that the distribution of $\omega$ differs among different dynamical subgroups; this has not been noticed previously because previous studies examined only the aggregate $\omega$ distribution of all the NEOs, e.g., \citet{Kostolansky:1999}.  Furthermore, we find that the apparent $\varpi$ distributions also differ among the dynamical subgroups.  We identify the origin of specific features in the observed non-uniform distributions with specific observational selection effects. Our analysis reveals several statistically significant non-random features in the intrinsic distributions and we identify specific dynamical effects that cause these features.

The rest of this paper is organized as follows.  In section~2, we introduce the NEO observational dataset that we have used and the definitions of the dynamical subgroups.  In section~3, 4 and 5, we analyze the distributions of the longitude of ascending node,  the argument of perihelion and the longitude of perihelion, respectively.  In section~6, we summarize our findings and conclusions.

\section{NEO observational data and dynamical subgroups}\label{s:definitions}
\begin{figure}
\centering
 \includegraphics[width=6in]{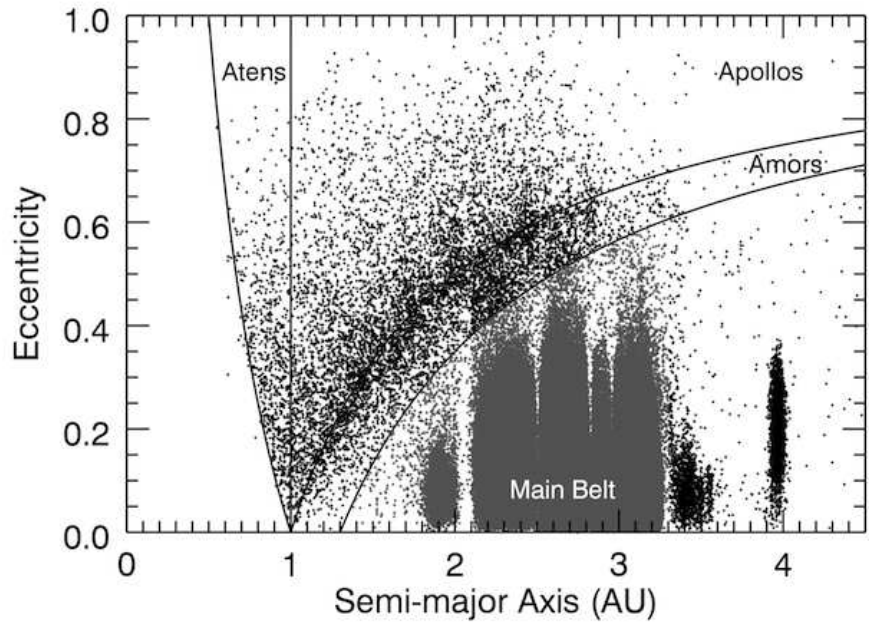}
 \caption{\sm Atens, Apollos, and Amors are delineated in the semimajor axis--eccentricity plane, as shown. The grey dots indicate the subset of main belt asteroids that are located within $a<3.3$~AU and $a(1-e)>1.3$~AU; the more distant main belt asteroids are not used in our analysis.}
\label{AstDistAE}
\end{figure}

We obtained the data on the NEOs from the Minor Planet Center (MPC), which maintains up-to-date lists of the minor planets of the solar system.  The dataset ``MPCORB" includes the well-determined osculating orbital elements and absolute brightness magnitude of 494613 objects, at epoch 10 April, 2013 in the J2000.0 ecliptic coordinate system. (We do not include objects that have poorly determined orbit solutions.) Of these, only 9590 objects meet the definition of NEO, i.e., perihelion distance $q<1.3$~AU.

We follow the MPC's definition of the NEO subgroups: 
\[
\begin{array}{ll}
\hbox{Atens}\quad & a< 1~\AU,\ a(1+e)> 1~\AU\\
\hbox{Apollos}\quad & a> 1~\AU,\ a(1-e)< 1~\AU\\
\hbox{Amors}\quad & 1~\AU <a(1-e)< 1.3~\AU
\end{array}
\]
where $a$ and $e$ denote the osculating semimajor axis and eccentricity.
Note that 19 Interior-Earth Objects (IEOs) which have aphelion distance smaller than 1~AU are excluded. The Atens, Apollos, and Amor groups consist of 747 (7.8\%), 4767 (49.8\%), and 4057 (42.4\%) members with median H magnitudes 22.6, 21.5, and 20.8, respectively.  We also make use of data of 475132 main belt asteroids that have $a<3.3$~AU and $a(1-e)>1.3$~AU; this sample is referred to as `MBAs' hereafter. The osculating semimajor axis vs.~eccentricity of all the minor planets in the inner solar system is shown in Figure \ref{AstDistAE}; the three NEO subgroups and the MBAs that are of interest in the present work are labeled in this diagram.

\begin{figure}
\centering
  \includegraphics[width=6in]{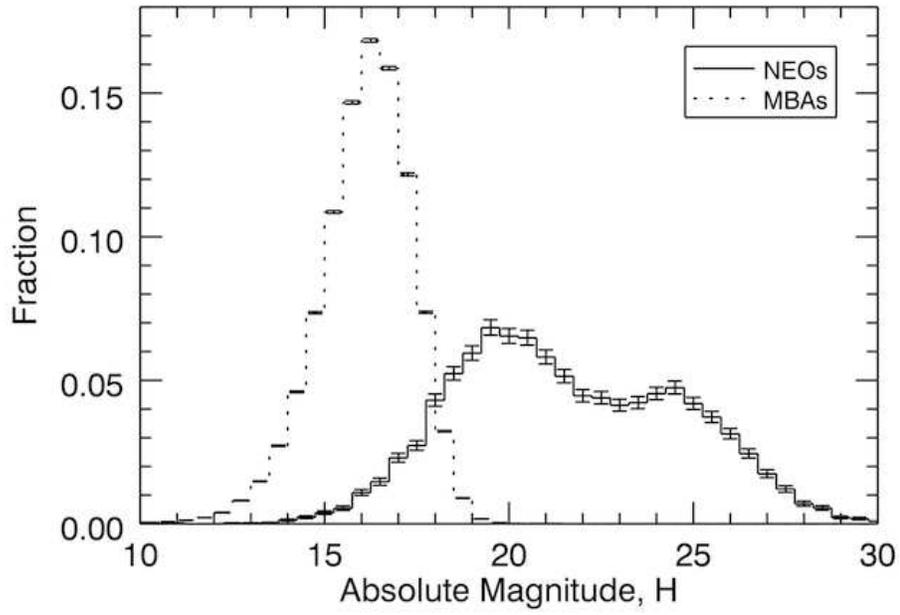}
 \caption{\sm The distribution of the absolute magnitude, $H$, of the NEOs and the MBAs.}
\label{DistHmag}
\end{figure}

In Fig.~\ref{DistHmag} we plot the distribution of the absolute magnitude, $H$, of the NEOs and MBAs.  (For all the figures in this paper, we adopt Poisson statistics, and plot error bars of $\pm\sqrt{N}$ where $N$ is the population in each bin for binned data.) As we can discern from these distributions,  the NEOs sample suffers from significant incompleteness for $H\gtrsim19$ and the MBAs are incomplete for $H\gtrsim15$.  We will make use of a subset of the NEO sample at faint magnitudes, $H>25$, to assess the most severe observational selection effects, as well as the subset of bright NEOs, $H<19$, as an approximation for an unbiased set of NEOs for the purpose of examining their intrinsic orbital distributions.  We will also make use of the subset of bright MBAs, $H<15$,  as an approximation for the observationally unbiased set of MBAs; the latter is considered observationally nearly complete~\citep{Tedesco:2005}.

\section{Longitude of Ascending Node}\label{s:nodes}
\begin{figure}
\centering
  \includegraphics[width=6in]{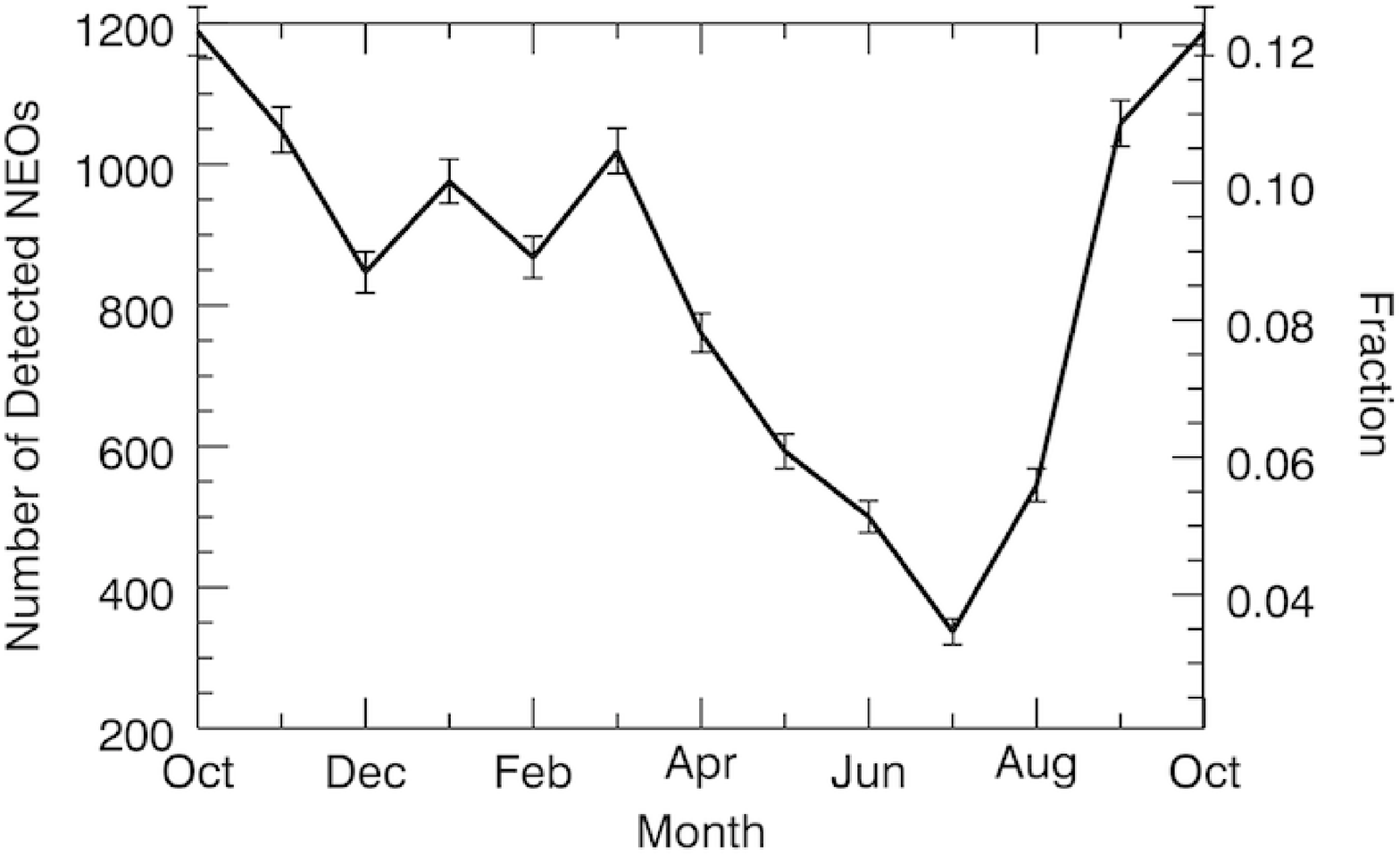}
 \caption{\sm Seasonal variation of all NEO discoveries (data up to April 10, 2013). Monthly bins are chosen to start from October because the September equinox ($\sim22$ September) is when the longitude of the opposition point is 0$^\circ$. Galactic plane crossing occurs during June and December.}
\label{AnnDetFreq}
\end{figure}

\begin{figure}
\centering
  \includegraphics[width=6in]{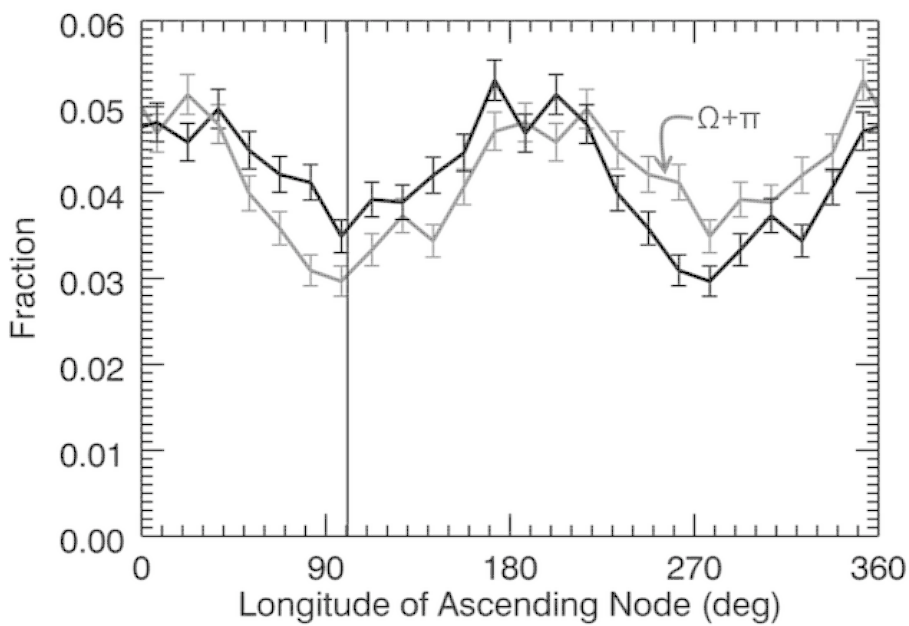}
 \caption{\sm Distribution of the longitude of ascending node, $\Omega$, of all NEOs (black line, total number N=9571). The gray line plots the same distribution but shifted by $\pi$ in the abscissa. The vertical line marks the location of Jupiter's longitude of ascending node, $\Omega_{J} = 100^\circ$.  The $\Omega$ distribution has systematically higher numbers around $\Omega_J$ compared to the opposite direction around $\Omega+180^\circ \simeq 290^\circ$). }                                  
\label{DistLNG_NEO}
\end{figure}
\begin{figure}
\centering
  \includegraphics[width=6in]{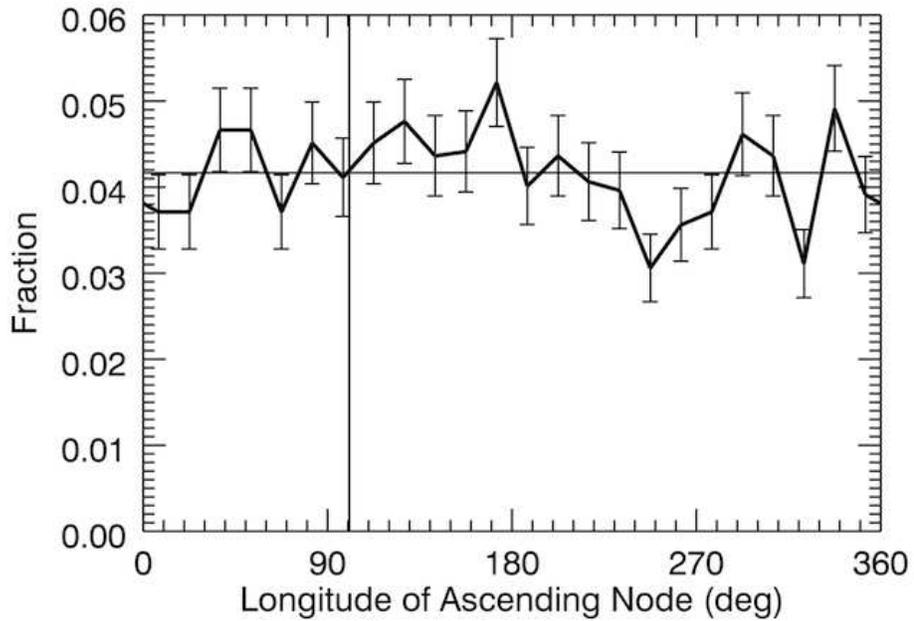}
 \caption{\sm Distribution of the longitude of ascending node, $\Omega$, of the bright NEOs ($H<19$, N=1994).  The mean direction of this distribution, $\bar\Omega=115^\circ$, is close to Jupiter's longitude of ascending node, $\Omega_{J} = 100^\circ$ (vertical line). The horizontal line indicates the mean fraction of $1/24$.}                                  
\label{DistLNG_NEO2}
\end{figure}

\begin{figure}
\centering
  \includegraphics[width=6in]{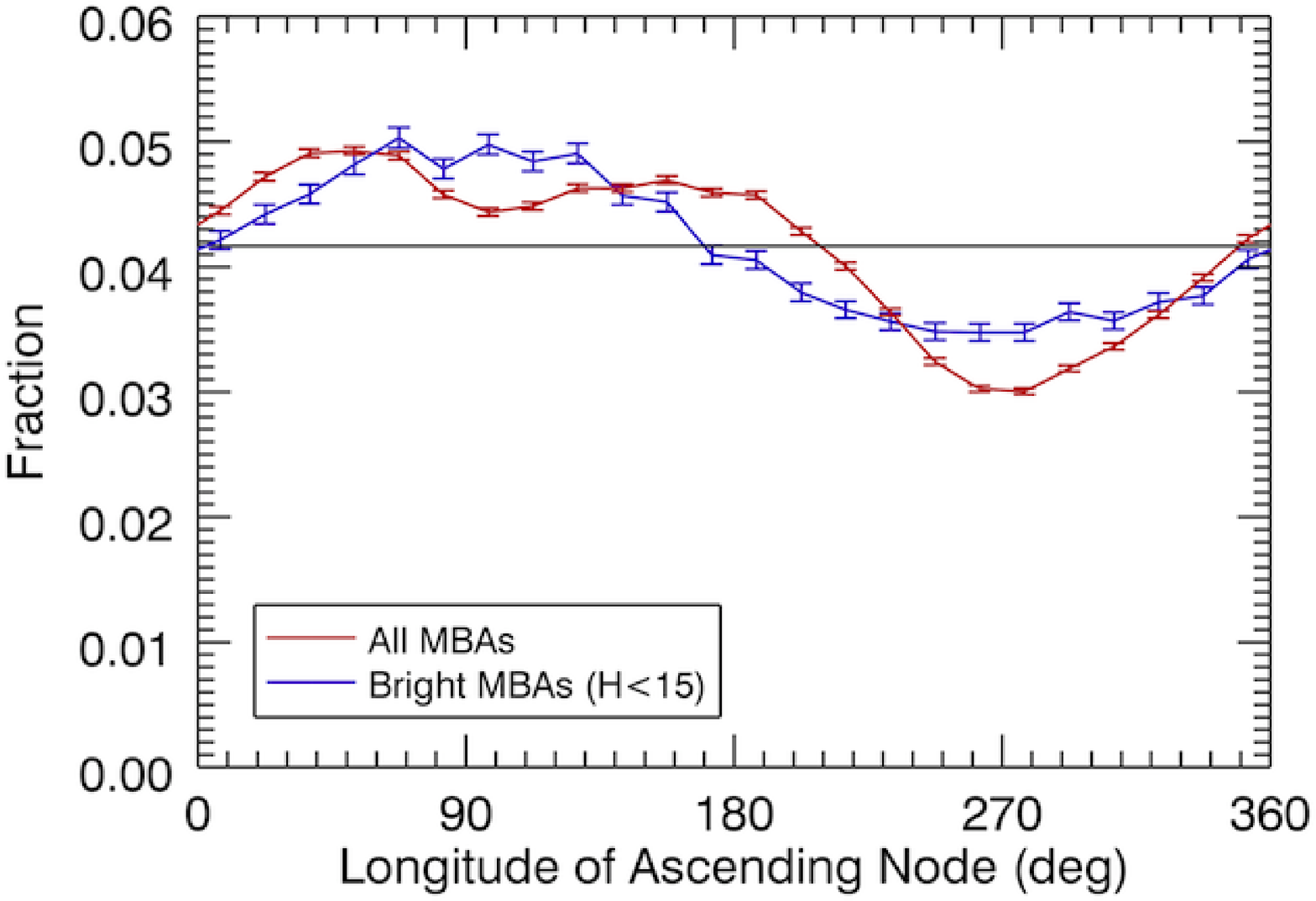}
 \caption{\sm Distribution of the longitude of ascending node, $\Omega$, of all MBAs (red line, N=475132), and of the bright MBAs ($H<15$, blue line, N=76889).}                                  
\label{DistLNG_MBA}
\end{figure}

To understand the sources of possible non-uniformities in the observed orbital distribution of NEOs, we first examine the annual variability in the discovery rate of asteroids. In Figure \ref{AnnDetFreq} we have plotted the number of NEO discoveries per month, as of April 10, 2013.  The monthly discovery rate varies by a factor of three, having the highest peak around October; the minimum discovery rate occurs in the month of July \citep{Emelyanenko:2011}.  

A similar annual variation also occurs for the MBAs.  \citet{Kresak:1989} explained these so-called `seasonal-effects' with three reasons. First, the Galactic plane is at opposition during June and December which adversely impacts the asteroid discovery rate due to the bright background. This adverse effect is more significant for the galactic center crossing during northern Summer than Winter.  Bad weather during Monsoon season in the southwestern United States, from early July to mid September, also reduces the discovery rate at the Catalina Sky Survey where a large fraction of discoveries have been made \citep{Jedicke:2002}. Secondly, longer nights during Winter in Northern hemisphere, where most discoveries are made, allow more observational time than during Northern Summer.  Thirdly, the concentration of MBAs' perihelion longitudes near 0$^\circ$ due to secular planetary perturbations enhances the discovery rate around September and October when opposition is in the direction of the perihelion of Jupiter. The non-uniform discovery rate of NEOs can be explained by similar reasoning. The first two effects are directly applicable to the NEOs' observations and the third one is relevant to the Amors but not to the Atens and Apollos; this is discussed in detail in section 5.

The seasonal variation of discovery rates can be expected to lead to a non-uniform distribution of $\Omega$.   Let us consider in some detail the expected non-uniformities.  Suppose Earth's longitude is $\lambda_{E}$ at a certain night. The observational bias for small geocentric distance and low phase angle favors the discovery of asteroids near $\lambda_{E}$ with low vertical distance from the ecliptic plane. Thus a high detection probability occurs for $\Omega=\lambda_{E}$ or $\Omega=\lambda_{E}+\pi$, respectively.  If we adopt the null hypothesis that the intrinsic $\Omega$ distribution is uniform, the number of discovered asteroids having $\Omega=\lambda_{E}$ and $\Omega=\lambda_{E}+\pi$ would be expected to be the same during that night. This argument holds for any observational night, even as observational conditions vary over time.   Consequently, the accumulated number of discovered asteroids having $\Omega \simeq \lambda_{E}$ and $\Omega \simeq \lambda_{E} + \pi$ should be indistinguishable and the $\Omega$ distribution of observed NEOs, $F(\Omega)$, should have a $\pi$ periodicity, i.e., $F(\Omega+\pi) = F(\Omega)$.

The $\Omega$ distribution of all NEOs is plotted with black line in Figure~\ref{DistLNG_NEO}.  (We adopt $15^\circ$ bins when we plot distributions of angular elements in this paper.) It is clear that the NEOs have a strongly non-uniform $\Omega$ distribution.  Moreover the distribution deviates from the $\pi$ periodicity expected from the above arguments: the range $0^\circ$ to $180^\circ$ contains 5021 NEOs, outnumbering by 471 the 4550 NEOs in the range $180^\circ$ to $360^\circ$; this is a $4.8\sigma$ departure from a random distribution (where $\sigma=\sqrt{N}/2$ is the standard deviation for binomial statistics).  The deviation from the $\pi$ periodicity is also apparent in the comparison of the black curve with the grey curve in Fig.~\ref{DistLNG_NEO}, as the latter plots the same $\Omega$ distribution but shifted by $\pi$ in the abscissa:  the black curve is systematically higher in the range 0--180$^\circ$ and systematically lower in the range $180^\circ$--$360^\circ$.
\citet{Kostolansky:1999} also previously noted the non-uniform $\Omega$ distribution of NEOs and briefly cited \citet{Kresak:1989} for explanation. But neither author explained the $\Omega$'s deviation from the expected $\pi$ periodicity.  

 In Figure \ref{DistLNG_MBA}, we  plot with red line the $\Omega$ distribution of our comparison set of MBAs. This also has a double dip pattern as in the NEOs, and a more pronounced deviation from the expected $\pi$ periodicity: there are 265750 MBAs in the range 0 to $180^\circ$ compared to 209382 in the range $180^\circ$ to $360^\circ$;  this corresponds to a $58\sigma$ departure from a random distribution (again, adopting binomial statistics).  It is tempting to directly relate the minimum in the $\Omega$ distribution near $270^\circ$ to the galactic center crossing in June, because the $\Omega$ distribution looks qualitatively similar to the annual discovery rate pattern (Figure \ref{AnnDetFreq}). However, for the reasons explained in the previous paragraph, the discovery rate variation causes non-uniformity with a $\pi$ periodicity, if the intrinsic $\Omega$ distribution were uniform. 
The deviation from the $\pi$ periodicity indicates that the intrinsic distribution may be non-uniform.

To determine the preferred intrinsic direction of $\Omega$, we calculate the mean angle, $\bar\Omega$, and its significance as follows.
The mean angle is defined as follows,
\begin{equation}
\sin\bar\Omega = {\sum_{i=1}^N \sin\Omega_i\over N}, \qquad
\cos\bar\Omega = {\sum_{i=1}^N \cos\Omega_i\over N},
\end{equation}
where $N$ is the sample size.  Then $r$, defined as follows,
\begin{equation}
r = \sqrt{\sin^2\bar\Omega + \cos^2\bar\Omega},
\end{equation}
lies in the range 0 to 1, and its value provides a measure of the dispersion in the data; a value $r=0$ indicates complete dispersion (no preferred direction) while a value $r=1$ indicates complete concentration in the direction of $\bar\Omega$.    The Rayleigh $z$-statistic, with the definition $z=Nr^2$, provides the statistical significance level of the directionality in angular data, by calculating the probability $p$ for the null hypothesis that the distribution is uniform around the circle~\citep{Fisher:1993}.  (The Rayleigh test assumes sampling from a von Mises distribution, a circular analog of the linear normal distribution.)  As usual, if the $p$--value is below a certain value $\alpha$, there is a probability $1-\alpha$ that the observed sample is non-uniform on the circle; $p<0.05$ is usually demanded for statistically significant results.  

Using the above procedure, we find $\bar\Omega=111^\circ$ and $r=0.041$ for the NEOs, and $\bar\Omega=94.5^\circ$ and $r=0.094$ for the MBAs.  Applying the Rayleigh test, we obtain  $p\ll10^{-3}$ for both the NEOs and the MBAs, indicating the high statistical significance of the directionality of the $\Omega$ distributions.  There are 5072 NEOs (53\%) having $\Omega$ within $\pm90^\circ$ of the peak direction, while  4499 (47\%) are outside this range; this is a $5.5\sigma$ departure from a uniform distribution (based on binomial statistics).  For the MBAs, the fractions are 56\% within $\pm90^\circ$ of the peak direction, and 44\% beyond that range, corresponding to an $82\sigma$ departure from a uniform random distribution.

For additional insight, it would be useful to examine an observationally unbiased sample of NEOs and MBAs.  For this purpose, we choose the subset of NEOs brighter than $H$ magnitude 19, and a subset of MBAs brighter than $H=15$. As noted in Section 2, these subsets are observationally nearly complete (cf.~Fig.~\ref{DistHmag}), and can be considered to approximate the intrinsic orbital distributions of the NEOs and the MBAs.   In Figure~\ref{DistLNG_NEO2}, we plot the $\Omega$ distribution of this subset of 1994 bright NEOs; the blue line in Figure~\ref{DistLNG_MBA} shows the $\Omega$ distribution of the 76889 bright MBAs.  We find $\bar\Omega=115^\circ$ and $r=0.038$ for the bright NEOs, and $\bar\Omega=89^\circ$ and $r=0.090$ for the bright MBAs.  Applying the Rayleigh test, we obtain $p=0.053$ and $p\ll10^{-3}$ for the bright NEOs and the bright MBAs, respectively.  In words, the bright MBAs have a statistically highly significant directionality of $\Omega$ but the significance of the directionality of $\Omega$ of the bright NEOs is marginal.   The latter is not surprising, as the relatively small number of bright NEOs yields a rather noisy $\Omega$ distribution. 

Given that the symmetry argument above indicates a statistically significant non-uniformity of the NEOs' $\Omega$ distribution, we proceed to consider its possible dynamical origins.
The excess of $\Omega$ values in the range $0^\circ$ to $180^\circ$ for main belt asteroids was already known a century ago and the reason was ascribed to the asteroids' mean plane departure from the ecliptic plane \citep{Plummer:1916}. The explanation is as follows.  If all asteroids revolved around the Sun in a common plane then all of them would have a single value of $\Omega$.  A dispersion about a mean plane would cause the $\Omega$ distribution to have a dispersion about the mean plane's longitude of ascending node.  \citet{Kresak:1967} noted that the mean plane of the first $\sim1660$ numbered asteroids (all the asteroids known at the time of Kresak's work) is close to the orbital plane of Jupiter; Jupiter's orbital plane is inclined $I_J=1.3^\circ$ to the ecliptic and has longitude of ascending node, $\Omega_J=100^\circ$~\citep{Murray:1999}.  This coincidence is explained by the effects of planetary perturbations as follows.  An asteroid's orbital plane can be described as the sum of a ``free" and a ``forced" inclination vector, where ``free" denotes a part that is set by (generally unknown) initial conditions (hence drawn from a random variate), and ``forced" denotes a part that is determined by planetary perturbations.  In linear secular theory, the latter depends only upon the semimajor axis of an asteroid~\citep{Murray:1999}.  Consequently, the mean plane of an ensemble of asteroids will coincide with the plane normal to the local forced inclination vector, $I_{f}(\cos\Omega_f,\sin\Omega_f)$.

\begin{figure}
\centering
  \includegraphics[width=6in]{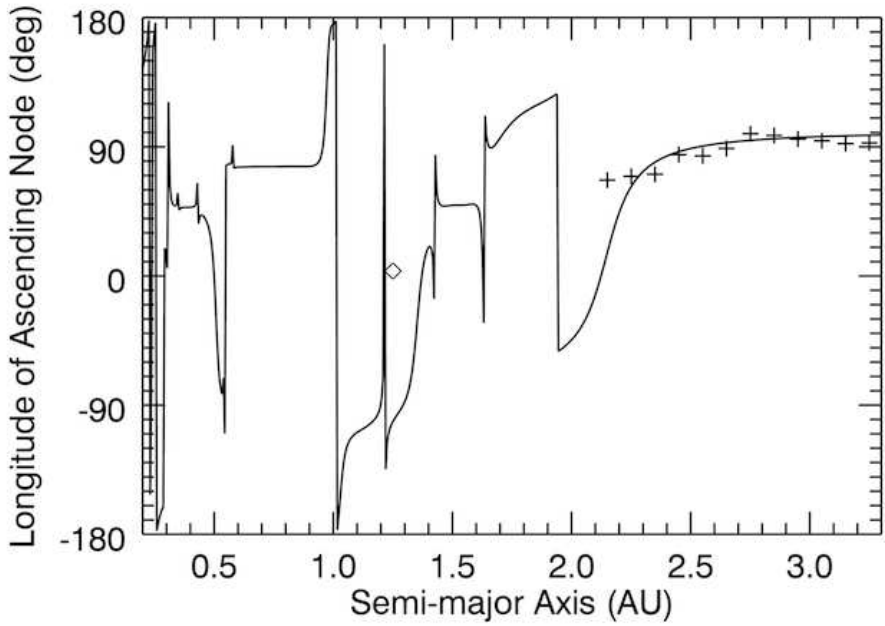}
 \caption{\sm The forced longitude of ascending node, $\Omega_{f} (a)$, as obtained from linear secular theory with eight planets, Mercury---Neptune is plotted as the black continuous line. Within each 0.1 AU semimajor axis bin, the mean directions, $\bar\Omega(a)$ of bright NEOs ($H<15$) are indicated with the `+' symbols if the directionality is statistically significant ($p<0.05$ for the Rayleigh z-test; see Table~\ref{TB_LNG}).  For those bins of low confidence level, we also calculated the peak directions for NEOs with $H<19$; only one bin, $1.2<a<1.3$~AU, is found to have statistically significant directionality, and the result is shown as the diamond symbol.}
\label{FrcdLNG}
\end{figure}

We calculated the forced inclination vector for the semimajor axis range of 0.2 AU to 3.3 AU using linear secular perturbation theory for the eight planets Mercury--Neptune \citep{Murray:1999}. 
In Figure \ref{FrcdLNG} we plot the forced value $\Omega_f$ as a function of the semimajor axis.  

We find that for the semimajor axis range $2$~AU~$<a<3.3$~AU, $\Omega_f$ is smoothly varying with $a$; for $a>2.4$~AU it has a nearly constant value near $90^\circ$.   For the semimajor axis range $2.1<a<2.4$~AU, $\Omega_f$ decreases as semimajor axis decreases and has values of $45^\circ$ and $-11^\circ$ at 2.2~AU and 2.1~AU, respectively.  
For $a\lesssim2$~AU, linear secular theory finds that the local forced plane varies rapidly with semimajor axis. This is due to the presence of the terrestrial planets as well as several secular resonances that cause the forced inclination vector to be very sensitive to the semimajor axis.  (Indeed, linear secular theory is likely insufficiently accurate to define the local forced plane for $a\lesssim2$~AU; a more accurate theory is beyond the scope of the present paper.)   

For each 0.1 AU size bin in semimajor axis, we calculated the mean values, $\bar\Omega (a)$, and the associated $p$--values for the bright minor planets ($H<15$)  using circular statistics as in equations (1)--(2); the results are given in Table \ref{TB_LNG}.  (Note that for this calculation, we do not distinguish between MBAs and NEOs.)  For $a<2.1$~AU, the sample size in each bin is too small to provide statistically significant results, but statistically significant directionality (with $p$--value $<0.05$) is found for the bins with $a>2.1$~AU.  For the latter range, the mean values, $\bar\Omega (a)$, are plotted as plus signs in Figure \ref{FrcdLNG}; we observe that they follow the theoretical curve quite well.  In the inner region of  $a<2.1$~AU, we  also carried out the analysis with a fainter magnitude cutoff ($H<19$), but found only one semimajor axis bin ($1.2<a<1.3$~AU) having statistically significant  ($p=0.04$) results for  directionality of $\Omega$; this is shown as the diamond symbol in Fig.~\ref{FrcdLNG}.  The small sample size and the rapid variation of the local forced plane for $a\lesssim2$~AU accounts for the poor directionality of the $\Omega$'s in this region.  However, integrated over the whole range of $a$ we find that the population of bright minor planets has a statistically significant concentration of $\Omega$ near $90^\circ$.  This accounts for the excess of $\Omega$ values in the range 0 to $180^\circ$ observed in the distribution of the MBAs (Fig.~\ref{DistLNG_MBA}) and possibly NEOs (Fig.~\ref{DistLNG_NEO}). 

In summary, the lack of $\pi$ periodicity and the high statistical significance of the directionality of the observed NEOs' $\Omega$ distribution cannot be explained by currently known observational biases.  The half-circle centered at about $\bar\Omega=111^\circ$ has 53\% of the NEOs and the complementary half-circle has only 47\%,  a 5.5$\sigma$ departure from a uniform random distribution. Thus, we cautiously claim that this is evidence of an underlying intrinsic non-uniformity caused by secular planetary perturbations, similar to that previously noted for the non-uniform $\Omega$ distribution of the MBAs.   The large variations of the local forced plane (over the semimajor axis range spanned by the NEOs), and the significant inclination dispersion of NEOs mutes their $\Omega$ directionality compared to that of the MBAs.   Consequently, direct confirmation of this result with an observationally complete NEO sample requires a sample size of at least $\sim2500$ for a $\sim3\sigma$ confidence level; the current sample of 1994 NEOs with $H<19$ falls slightly short. We predict that the intrinsic non-uniform $\Omega$ distribution of the NEOs will be further revealed with a larger observationally complete sample size.

\begin{table}
\centering
\caption{Mean directions of the longitudes of ascending node, $\bar\Omega (a)$, of minor bodies and their significance probabilities, $p$. The bright minor bodies (absolute magnitude $H<15$) are binned in 0.1~AU bins in semimajor axis, $a$, between 1.7 and 3.3~AU. For the bins $a<1.7$~AU, the sample sizes are exceedingly small, so we used a fainter magnitude cut-off ($H<19$) and tabulated here if they met the condition of $p<0.05$. }
\begin{tabular}{| l | p{2cm} | l | l | l |}
\hline
  Bin & absolute magnitude cut-off& Number & Mean angle (deg) & $p$\\
  \hline  \hline
    $1.2<a<1.3$~AU & 19 & 73 & 3.4 & 4.03$\times 10^{-2}$ \\ 
  $1.7<a<1.8$~AU & 15 & 10 & -80.1 & 5.84$\times 10^{-2}$ \\
  $1.8<a<1.9$~AU & 15 & 86 &  63.0 & 5.30$\times 10^{-1}$\\
  $1.9<a<2.0$~AU & 15 & 273 & 57.0 & 8.66$\times 10^{-1}$\\
  $2.0<a<2.1$~AU & 15 & 9 & 147.1 & 9.97$\times 10^{-1}$\\
  $2.1<a<2.2$~AU & 15 & 787 & 66.7 & 1.73$\times 10^{-3}$\\
  $2.2<a<2.3$~AU & 15 & 4147 & 69.4 & $< 10^{-3}$\\
  $2.3<a<2.4$~AU & 15 & 5593 & 70.7 & $< 10^{-3}$\\
  $2.4<a<2.5$~AU & 15 & 3805 & 84.5 & $< 10^{-3}$\\
  $2.5<a<2.6$~AU & 15 & 7113 & 83.6 & $< 10^{-3}$\\
  $2.6<a<2.7$~AU & 15 & 9370 & 88.7 & $< 10^{-3}$\\
  $2.7<a<2.8$~AU & 15 & 7618 & 99.0 & $< 10^{-3}$\\
  $2.8<a<2.9$~AU & 15 & 3029 & 97.8 & $< 10^{-3}$\\
  $2.9<a<3.0$~AU & 15 & 5618 & 95.3 & $< 10^{-3}$\\
  $3.0<a<3.1$~AU & 15 & 10681 & 94.2 & $< 10^{-3}$\\
  $3.1<a<3.2$~AU & 15 & 15016 & 92.1 & $< 10^{-3}$\\
  $3.2<a<3.3$~AU & 15 & 3772 & 92.7 & $< 10^{-3}$\\
\hline 
\end{tabular}
\label{TB_LNG}
\end{table}

\section{Argument of Perihelion}\label{s:peri}
\begin{figure}
\centering
   \includegraphics[width=6in]{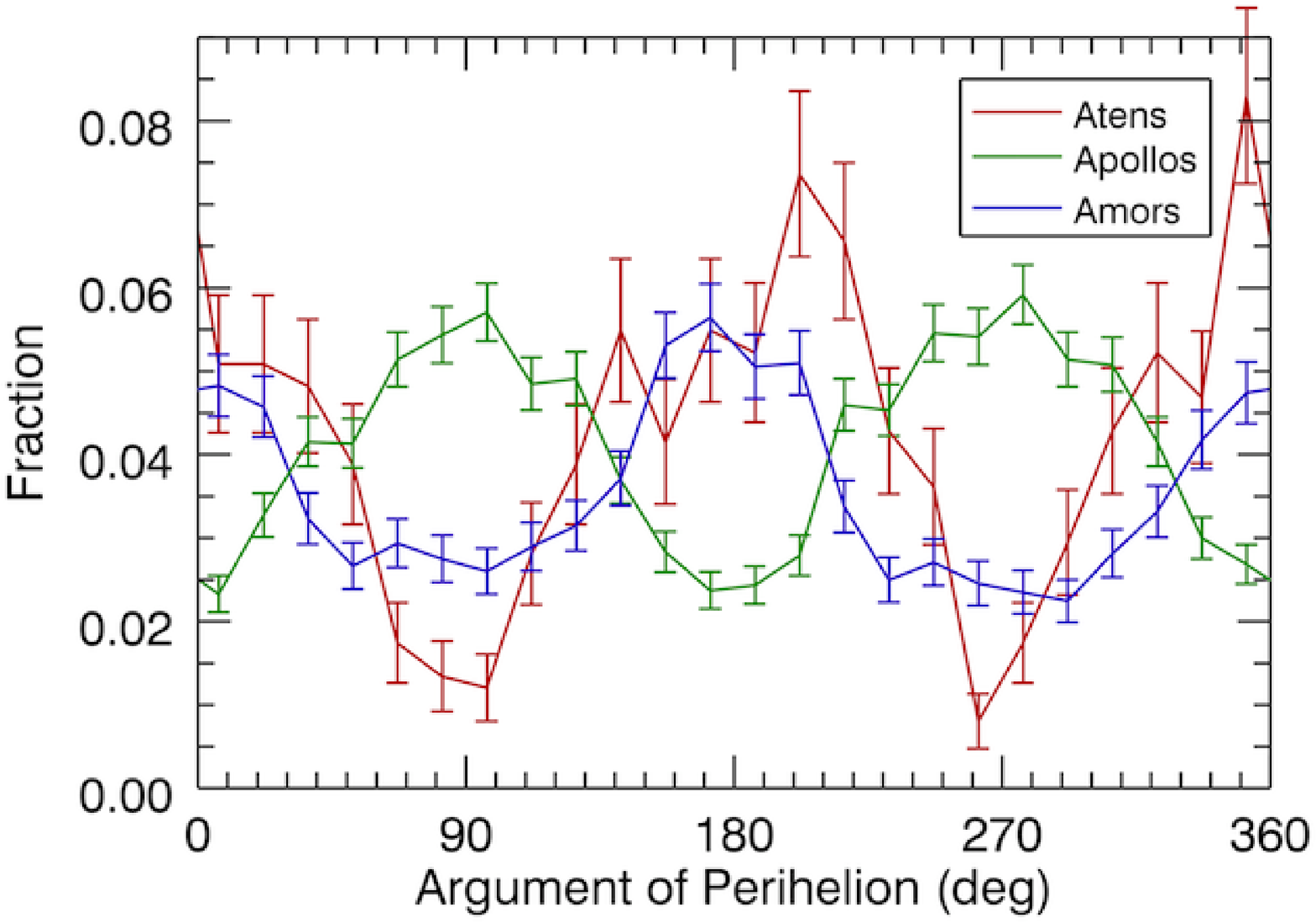}
  \caption{\sm The distributions of the argument of perihelion, $\omega$, for 747 Atens (red), 4767 Apollos (green), and 4057 Amors (blue).}
\label{DistAGPNEOs}
\end{figure}

We plot in Figure \ref{DistAGPNEOs} the distributions of the argument of perihelion, $\omega$, of the three subgroups of NEOs.  We observe that each subgroup has a different and distinct distribution:  the innermost group, Atens, and the outermost group, Amors, avoids $\omega \simeq 90^\circ$ and $\omega \simeq 270^\circ$, whereas the intermediate group, Apollos, favors these regions. These peculiar patterns in the distributions have not been noted previously. \citet{Kostolansky:1999} plotted the $\omega$ distribution but incorrectly concluded that the observed $\omega$ values are uniformly distributed because the author combined all the subgroups; the aggregate sample hides the striking patterns that exist in the subgroups. We show in this section that most of these peculiarities are due to observational bias, but planetary perturbations also have an effect. 

\begin{figure}
\centering
   \includegraphics[width=6in]{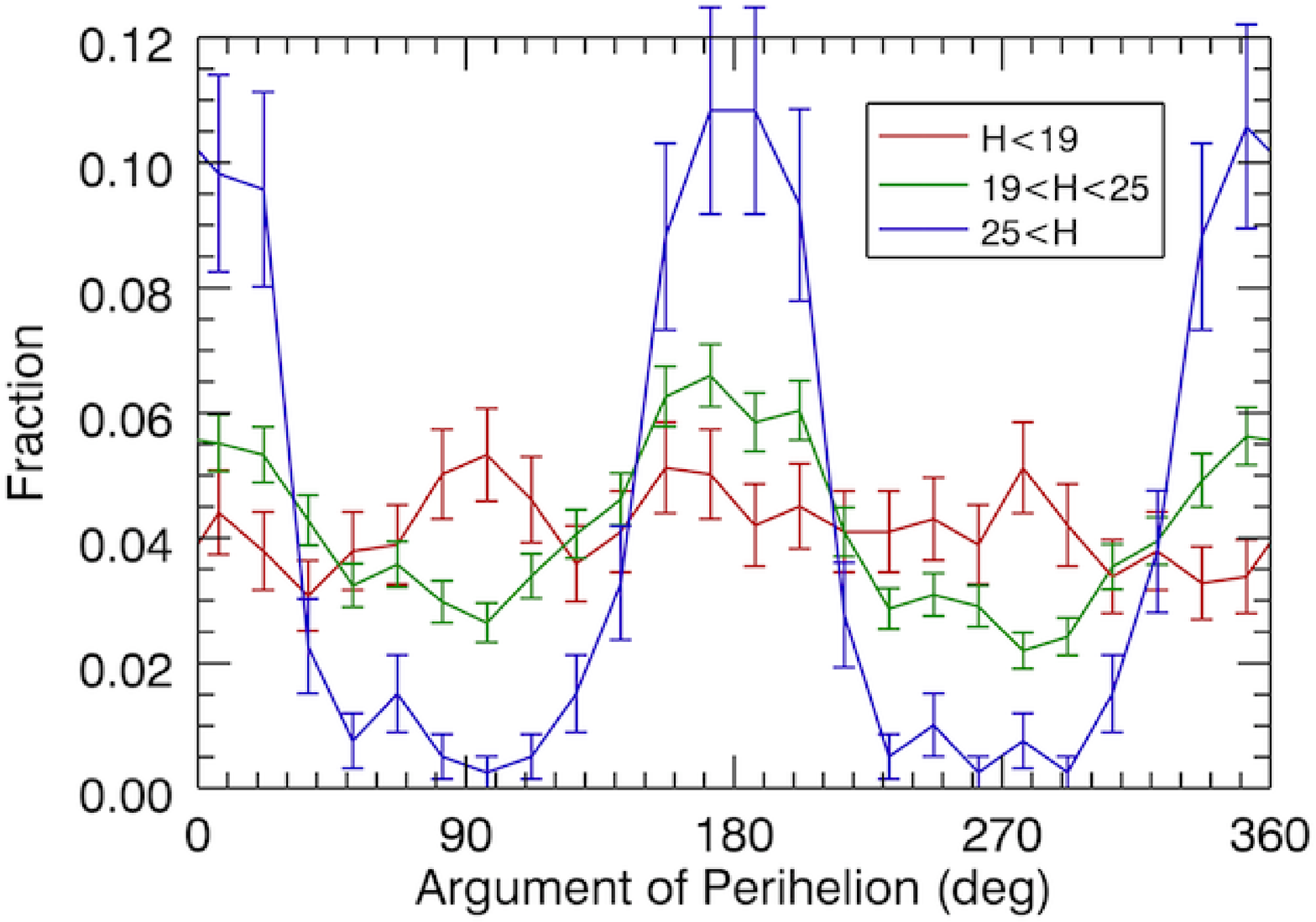}
  \caption{\sm The distributions of the argument of perihelion, $\omega$, for the bright Amors ($H<19$, red line, N=976), intermediate brightness Amors ($19<H<25$, green line, N=2684), and faint Amors ($H>25$, blue line, N=397), respectively.}
\label{DistAmorAGP2025}
\end{figure}

Amors are the outermost subgroup of NEOs; they do not cross Earth's orbit but have perihelia just beyond 1 AU. The Amors have high eccentricities, with a mean value of 0.41.  About 70 percent of Amors have aphelion distance more than twice farther than their perihelion distance. Thus, in magnitude-limited sky surveys, the detection of faint Amors is limited to the time when they are passing perihelion. Additionally,  close proximity to Earth requires low vertical distance from the ecliptic plane. This favors the detection of objects whose perihelion is located near their node, i.e. $\omega \simeq 0$ or $\omega \simeq 180^\circ$. In Figure \ref{DistAmorAGP2025} we show three plots of the $\omega$ distribution of subgroups of Amors of different brightness: bright ($H<19$), intermediate ($19<H<25$), and faint ($H>25$) objects. From this plot, we note that very few faint Amors (blue line) with $\omega \simeq 90^\circ$ or $270^\circ$ have been observed, which is consistent with expectations of observational selection effects.  In contrast, the bright Amors (red line) have an almost uniform random $\omega$ distribution: the Rayleigh z--test gives $p=0.062$.  We also applied the Rayleigh $z$--test to the distribution of the double angle, $2\omega$, and found no statistical significance ($p=0.44$) for any axial preference. These results indicate that the intrinsic $\omega$ distribution of the Amors is consistent with uniform random. 

\begin{figure}
\centering
   \includegraphics[width=6in]{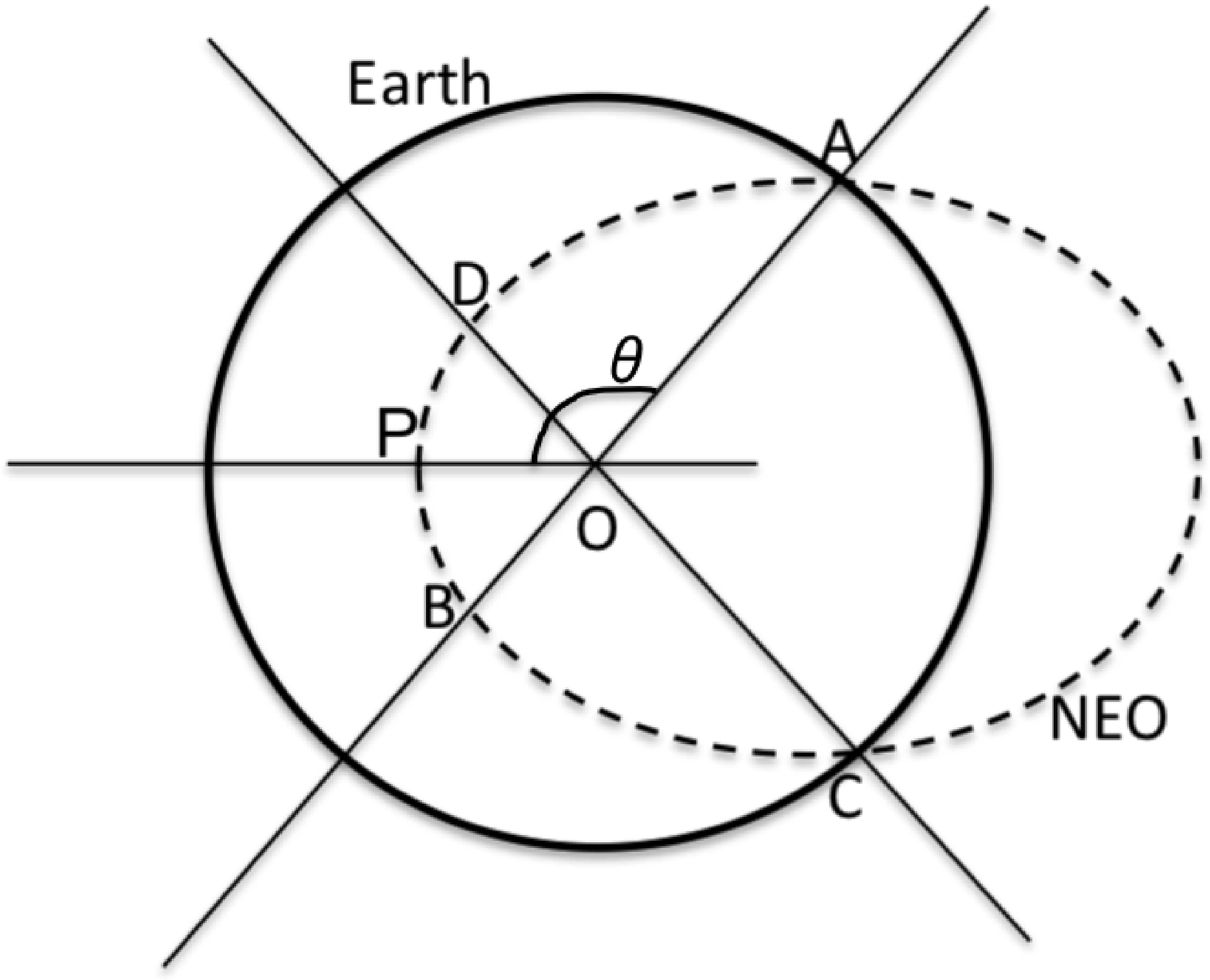}
  \caption{\sm Schematic diagram of the orbits of Earth and an Earth-crossing NEO, projected on the ecliptic plane. Two possible locations of nodal lines are denoted as AB and CD.  The most favorable observational condition for a faint NEO is when the NEO is at its nodal crossing and just outside Earth's orbit with heliocentric distance $\sim1$~AU, i.e., either at point A or at point C.}
\label{diagram}
\end{figure}

\begin{figure}
\centering
\includegraphics[width=220px]{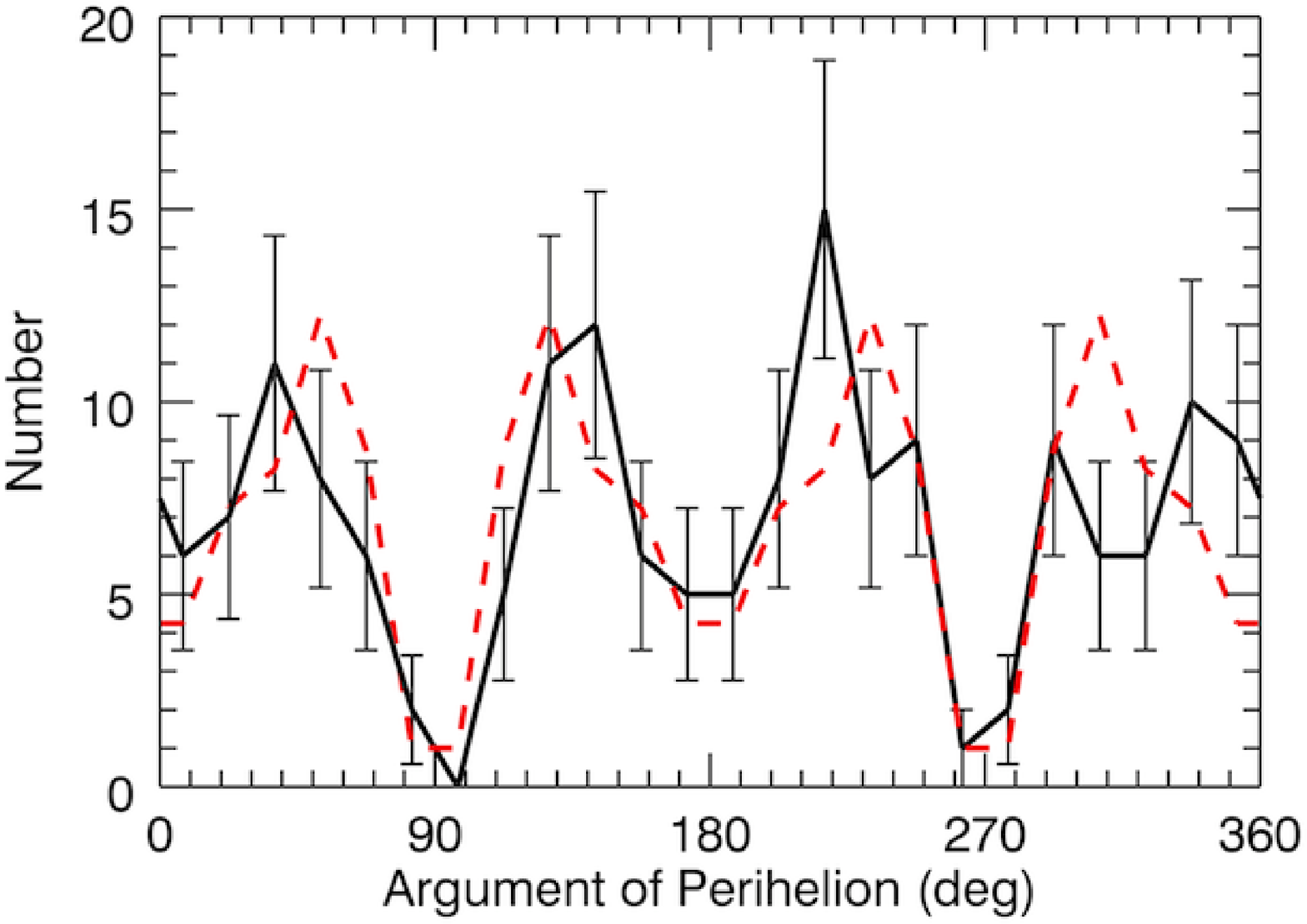}
\includegraphics[width=220px]{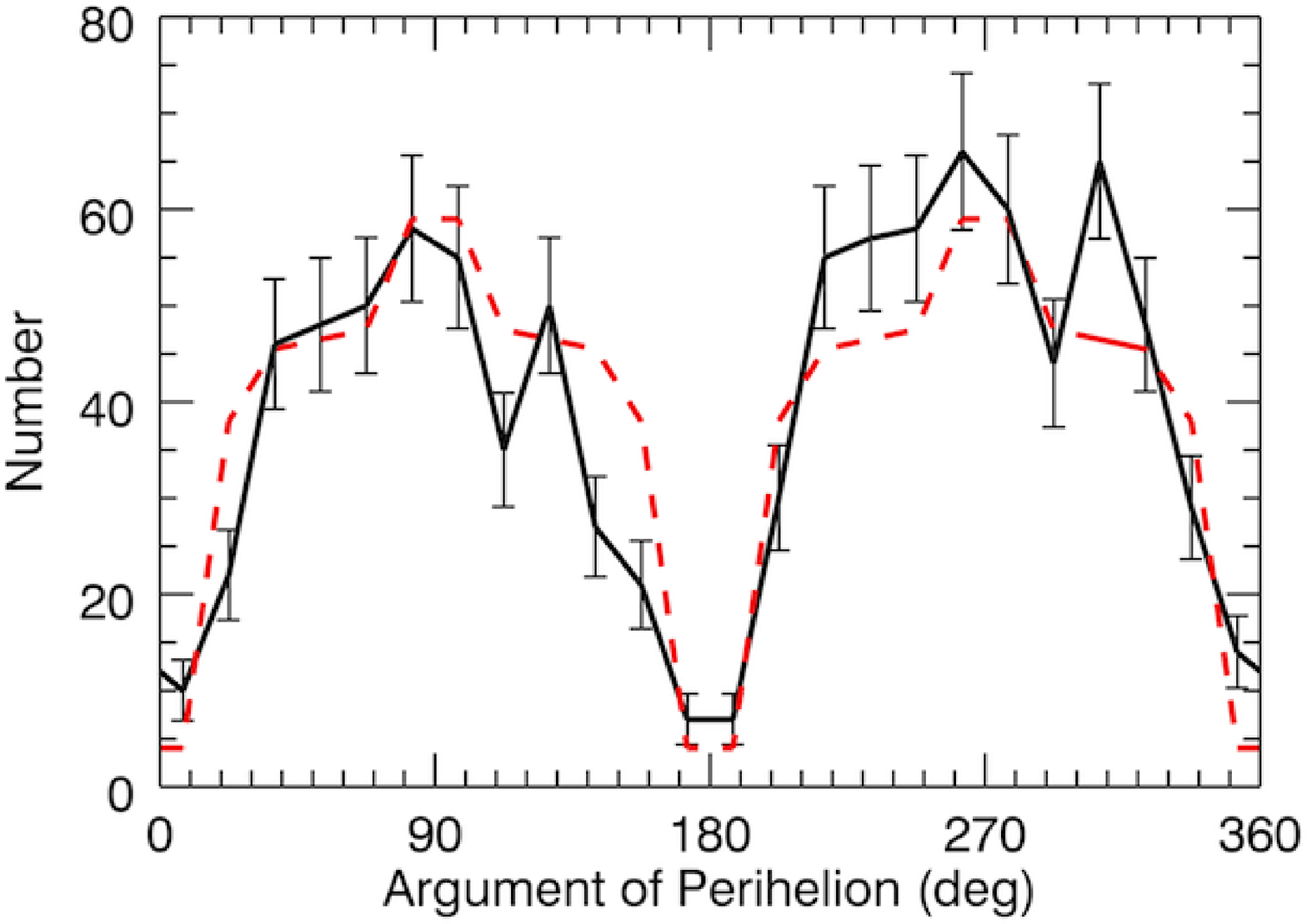}
  \caption{\sm Distribution of the argument of perihelion, $\omega$, for faint Atens (left panel, N=167) and for faint Apollos (right panel, N=962). Only faint objects ($H>25$) are included to clearly reveal the observational bias. The black solid line denotes the observed distribution of $\omega$ and the red dashed line denotes the theoretical distribution that models the observational bias. See Section~\ref{s:peri} for explanation. }
\label{DistECAo25}
\end{figure}

Atens and Apollos are objects whose orbits are Earth-crossing. A schematic diagram of their orbital shape is shown in Figure \ref{diagram}.  Faint objects are most observable when they approach close to Earth and present a low phase angle, i.e., when these NEOs are close to but just a little bit exterior to Earth's heliocentric location.  In this geometry, they are near the ecliptic plane and have a heliocentric distance $\sim1$~AU. Thus, the nodal line of observationally favored faint NEOs is along either the AB line or the CD line in Figure~\ref{diagram}, and detection is favored when these NEOs are located either at point A or at point C. In these favored geometries, the argument of perihelion, $\omega$, is related to the angle AOP, $\theta$, and to the true anomaly, $\vartheta$, as follows:
\begin{equation} \label{eq:OmgTh}
\omega = \left\{
\begin{array}{lll}
\theta &=2\pi-\vartheta &  \hbox{if A is the ascending node,}\\
\pi + \theta &= \pi - \vartheta & \hbox{if A is the descending node,}\\
2\pi - \theta &= 2\pi - \vartheta  &\hbox{if C is the ascending node,}\\
\pi - \theta &= \pi - \vartheta & \hbox{if C is the descending node.}
\end{array}\right .
\end{equation}
The angle $\theta (a,e)$ can be calculated with the assumption that the distance AO~$= 1+\varepsilon$~AU:
\begin{eqnarray} \label{eq:AO1AU}
\frac{a (1-e^2)}{1+e \cos \theta} = 1+\varepsilon\quad\implies\quad \theta(a,e,\varepsilon) = \arccos\,[(a(1-e^2)/(1+\varepsilon) -1)/e].
\end{eqnarray}
We generate synthetic distributions of the argument of perihelion subject to the above observability constraint, as follows. We adopt the $(a,e)$ values of the Atens and Apollos, respectively.  We choose a random value of $\varepsilon$ in the range $(0.0,0.02)$~AU.  Then, each combination of $(a,e,\varepsilon)$,  $\theta (a,e,\varepsilon)$ yields four possible values of the longitude of perihelion, $\omega$~(Eq.~\ref{eq:OmgTh}); we assume that these four solutions are equally probable. (The median difference between the observed $\omega$ and the nearest calculated value of $\omega'$ was small, $6.0^\circ$ and $4.5^\circ$, respectively.)  In Figure \ref{DistECAo25}, we plot the synthetic distributions as well as the observed distributions for the faint Atens and Apollos (H $>$ 25).  In the case of the Apollos, two peaks are dominant near $90^\circ$ and $270^\circ$. In contrast, in the case of Atens, the lowest points are near $90^\circ$ and $270^\circ$ and there are relatively shallower minima  near $0$ and $180^\circ$.  We also mention that brighter Atens have even higher populations for $\omega$ near $0$ and $180^\circ$ (not shown in the figure). The brighter Atens ($H<25$) remain observable at larger heliocentric distance than 1.02~AU; this explains their relatively higher rate of discovery near $\omega \simeq 0$ and $\omega \simeq 180^\circ$, respectively~(cf.~Fig.~\ref{DistAGPNEOs}).  The  agreement between the observed and the theoretical distributions demonstrates that the observational bias due to their orbital geometry explains the non-uniform distribution of the argument of perihelion of the faint Atens and Apollos.

\begin{figure}
\centering
   \includegraphics[width=6in]{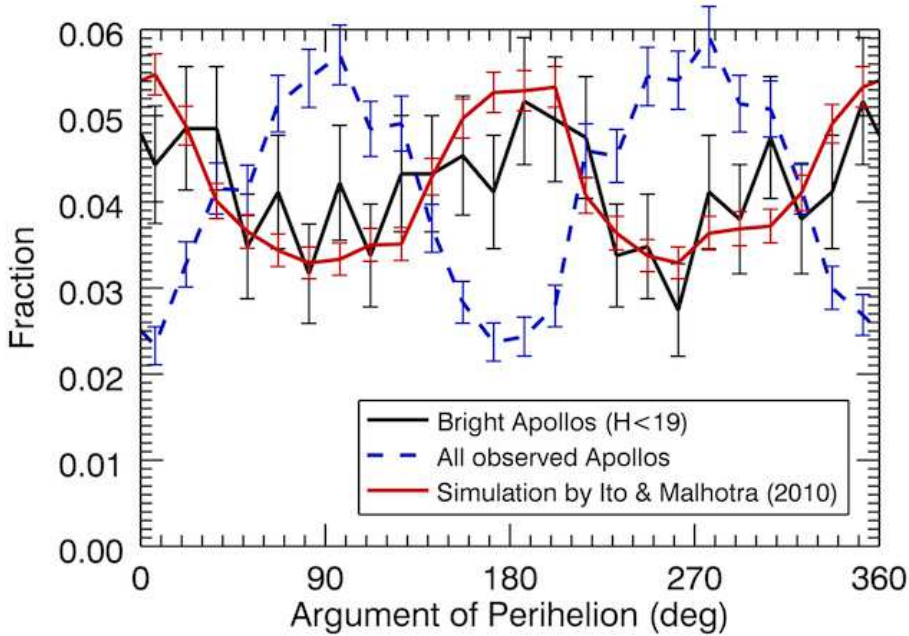}
  \caption{\sm Distribution of the argument of perihelion, $\omega$, for Apollos. The solid black line denotes bright Apollos ($H<19$, N=948) while the dashed blue line represents all the Apollos (N=4767). The red solid line is a simulated $\omega$ distribution of Apollos (N=9602) from \citet{Ito:2010}. }
\label{AGP_Apol}
\end{figure}

\begin{figure}[h]
\centering
\includegraphics[width=220px]{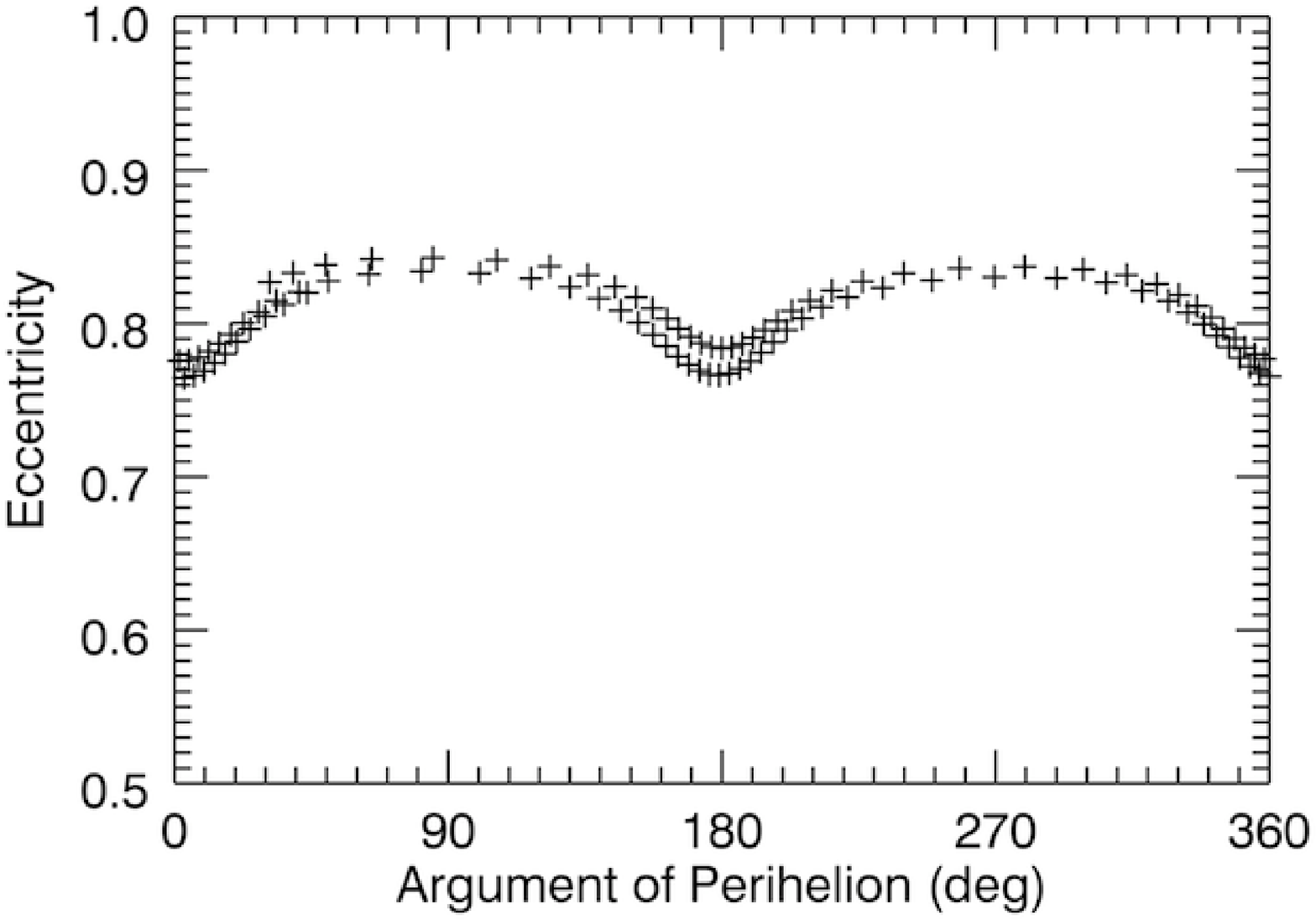}
\includegraphics[width=220px]{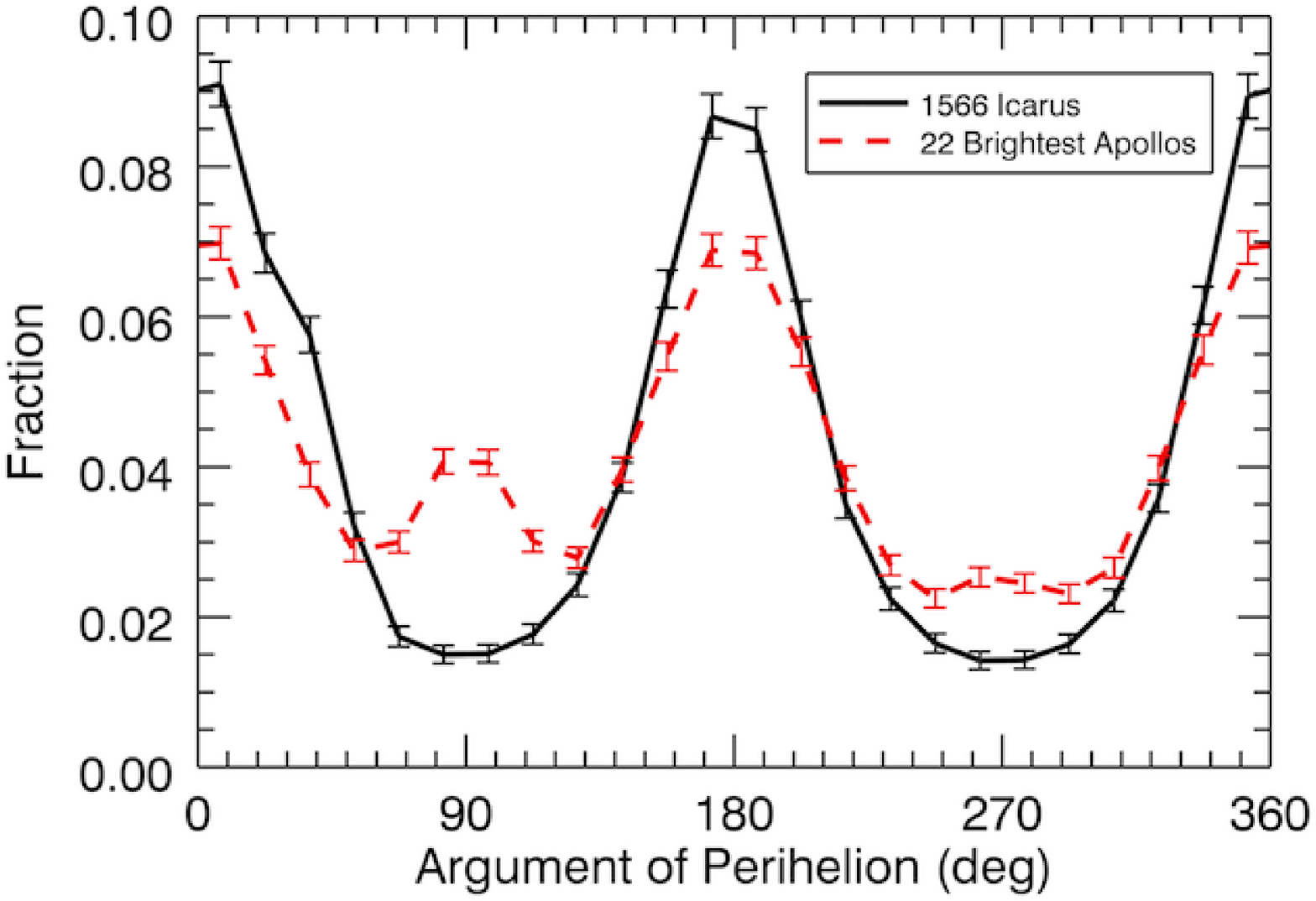}
\caption{\sm Dynamical evolution of an Apollo group NEO, 1566 Icarus, for $10^{5}$ year with initial conditions taken at J2000.0. Its argument of perihelion circulates with a strongly non-uniform rate (left panel); one data point is plotted every thousand years to reveal the non-uniform time-variation in phase space. The time-integrated $\omega$ distribution is plotted in the right panel (black curve), which clearly shows 1566 Icarus spends more time around 0 and $\pi$.  The $10^6$ years time-integrated $\omega$ distribution of the 22 brightest Apollos ($H<15$) is plotted as the red dashed curve.}
\label{Icarus}
\end{figure}

To discern the intrinsic distribution of $\omega$ of these subgroups, we turn to the subset of bright objects.   In the case of bright Atens ($H<19$), their number is relatively small ($N=70$), and the Rayleigh z--test finds no statistically significant directionality.  However, the bright Apollos' $\omega$ distribution is more interesting.  In Figure~\ref{AGP_Apol} we plot the $\omega$ distribution of bright Apollos ($H<19$, N=948) to compare with the distribution for all Apollos. The latter distribution is similar to that of the faint Apollos shown in Figure~\ref{DistECAo25} as the faint objects $(H\ge19)$ are numerically dominant, comprising approximately 80\% of the total Apollo population.  Interestingly, the bright Apollos have the opposite trend compared to the distribution of the full sample, with peaks near $\omega\simeq 0$ and $\omega=180^\circ$.  This distribution appears to be axial.  With $\omega$ defined to be in the range $-180^\circ$ to $+180^\circ$, we find that there are 522 bright Apollos having $|2\omega|<90^\circ$ and only 426 in the complementary range. This is a $3.2\sigma$ departure from a uniform random distribution.  We also applied the Rayleigh z--test to the distribution of the double angle, $2\omega$.  We find that $2\omega$ has a mean value of $3^\circ$, with a $p$--value of $5.6\times 10^{-3}$, indicating high statistical significance of the directionality of this distribution.  We conclude that the bright Apollos have an axial distribution of $\omega$, with double peaks at $0$ and $180^\circ$; fifty-five percent of Apollos have $\omega$ in the two quadrants centered at these peak directions.  This non-uniform feature is indicative of underlying dynamical effects.

We also turned to a large synthetic dataset of Apollos that was available to us from a dynamical simulation of NEOs~\citep{Ito:2010}.  In this simulation, the authors generated a steady state orbital distribution of NEOs by numerically integrating the orbits of test particles having initial orbital distribution (in $a,e,i$) the same as the de-biased distribution for the NEOs with $H<18$ computed by~\citet{Bottke:2002} and adopting uniform random initial angular orbital elements; the test particle orbits were integrated under all planets' perturbations; particles that collided with the Sun or a planet were replaced with identical source initial conditions.  (The simulation is admittedly imperfect in several ways; for example, it does not employ a detailed physical model of the source regions of NEOs, rather it re-generates most frequently the most unstable orbits.  Nevertheless, we can treat this simulation as a numerical experiment to obtain insights into the NEO dynamics.)  To our surprise, we found that in this simulation, the $\omega$ distribution of the 9602 simulated Apollos (shown as the red solid curve in Figure~\ref{AGP_Apol}) is similar to the observed bright Apollos'.  The population per bin varies by more than 50\% from the minima at $\pm90^\circ$ to the maxima at $0$ and $180^\circ$. Because the simulation re-generates lost (unstable) particles more frequently and with a uniform random distribution of angular elments, it tends to suppress any dynamically-induced non-uniformity; this suggests that the intrinsic non-uniformity might be even greater.  This supports the conclusion that the observed axial pattern of the bright Apollos is real, and indicates that it is caused by dynamical effects.

To understand the axial pattern in the $\omega$ distribution of the Apollos, we note that the perihelion and aphelion of Apollos are both located away from 1 AU. Thus, Apollo asteroids having non-zero inclination and $\omega\simeq 0$ or $\omega=\pi$ reach 1~AU heliocentric distance off the ecliptic plane. This configuration prevents very close encounters with Earth, and thereby favors longer dynamical lifetimes.  A long-known and exemplary Apollo asteroid, 1566 Icarus, illustrates this effect well.  This asteroid exhibits coupled variations of inclination--eccentricity--$\omega$.  We carried out a long-term (100 kyr) integration of 1566 Icarus' orbit. We used the symplectic orbit integration package SWIFT--RMVS3 (\url{http://www.boulder.swri.edu/~hal/swift.html}); we included the perturbations from all the major planets, and we used a step size of 1 day to resolve accurately the small perihelion distances of these objects.   In Figure~\ref{Icarus}, in the left panel, we plot the evolution of 1566 Icarus in the eccentricity--$\omega$ plane; one point is plotted every 1000 years.  We see that the argument of perihelion circulates with a period of about 50 kyr, consistent with a previous study of 1566 Icarus~\citep{Ohtsuka:2007}.  During the circulation of $\omega$, it spends more time around $\omega = 0$ and $\omega=\pi$.  The time-averaged distribution of $\omega$ for 1566 Icarus is plotted as the solid black curve in the right panel of Figure~\ref{Icarus}; we observe that it is qualitatively similar to the $\omega$ distribution of bright Apollos.  To explore further,  we also carried out long-term (1 Myr) integrations of the orbits of the 22 brightest Apollo asteroids ($H<15$); during this time span, some of these objects transfer into other subgroups as their semimajor axes and eccentricities evolve, and their $\omega$s occasionally change behavior from circulation to libration and vice versa.  During the time when they appear in the Apollo dynamical  group, we calculated their time-averaged distribution of $\omega$; this is shown as the red dashed curve in the right panel of Figure~\ref{Icarus}.  This striking non-uniform distribution is clearly caused by secular planetary perturbations.   However, the amplitude of this modelled non-uniform distribution is much larger than that of the observed population of bright Apollos (cf.~Figure~\ref{AGP_Apol}).

The dynamical libration of $\omega$ about $90^\circ$ or $270^\circ$ has been known to occur for highly inclined orbits due to the distant perturbations of Jupiter \citep{Kozai:1962}. (The difference between the ecliptic plane and the plane of Jupiter's orbit makes only a small difference to the centers of $\omega$ libration in this context.)  Circulations of $\omega$ under the so-called Kozai effect can be expected to yield a distribution qualitatively similar to that observed in the right panel of Figure~\ref{Icarus}. For small amplitude librations, the time-averaged distribution of $\omega$ would peak near the centers of libration at $\pm90^\circ$; however, large amplitude librations would lead to a time averaged $\omega$ distribution having peaks away from the libration center because objects spend more time near the extremes of libration.   Thus, a combination of circulating and librating behavior under secular planetary perturbations is the likely underlying cause of the non-uniform $\omega$ distribution in Figure~\ref{AGP_Apol}.   Further investigation of this interesting dynamics is warranted.

\section{Longitude of Perihelion}\label{s:varpi}
\begin{figure}
\centering
 \includegraphics[width=6in]{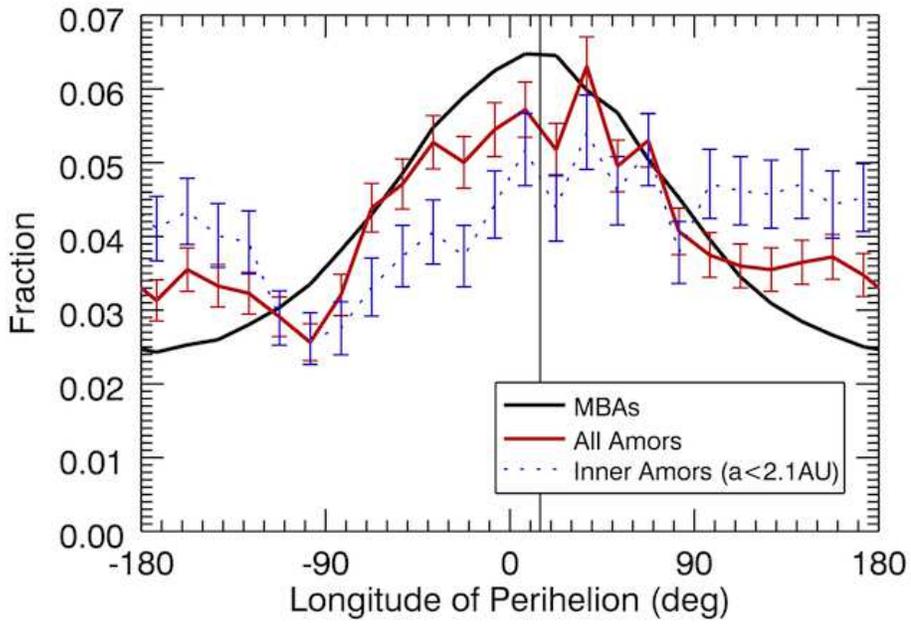}
 \caption{\sm Distribution of Amors' longitude of perihelion, $\varpi$ (red solid line). The perihelion longitude distribution of all main belt asteroids is also plotted (black solid line). The blue dotted line shows the distribution for the subset of Amors having semimajor axis $a < 2.1~\AU$ (N=2144). The vertical line indicates the longitude of perihelion of Jupiter, $\varpi_J=15^\circ$.}
\label{DistLNP}
\end{figure}

\begin{figure}
\centering
 \includegraphics[width=6in]{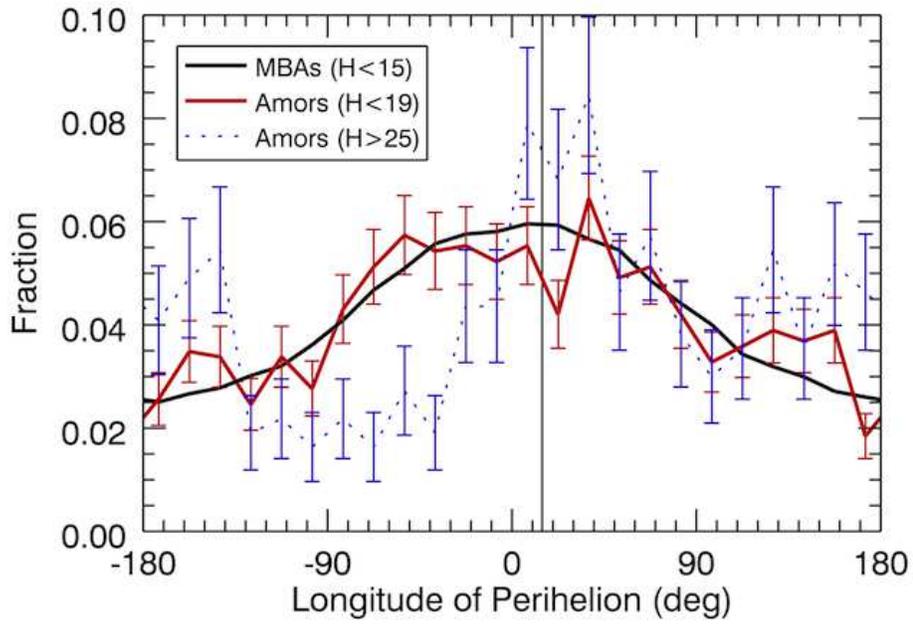}
 \caption{\sm Distribution of Amors' perihelion longitude $\varpi$ for the bright objects ($H<19$, N=976, red solid line) and the faint ones (H$>25$, N=367, blue dotted line). For comparison, we also plot the $\varpi$ distribution of the bright MBAs ($H<15$, N=76889, black solid line). The vertical line indicates the longitude of perihelion of Jupiter, $\varpi_J=15^\circ$. }
\label{DistLNP2}
\end{figure}

\begin{figure}
\centering
 \includegraphics[width=6in]{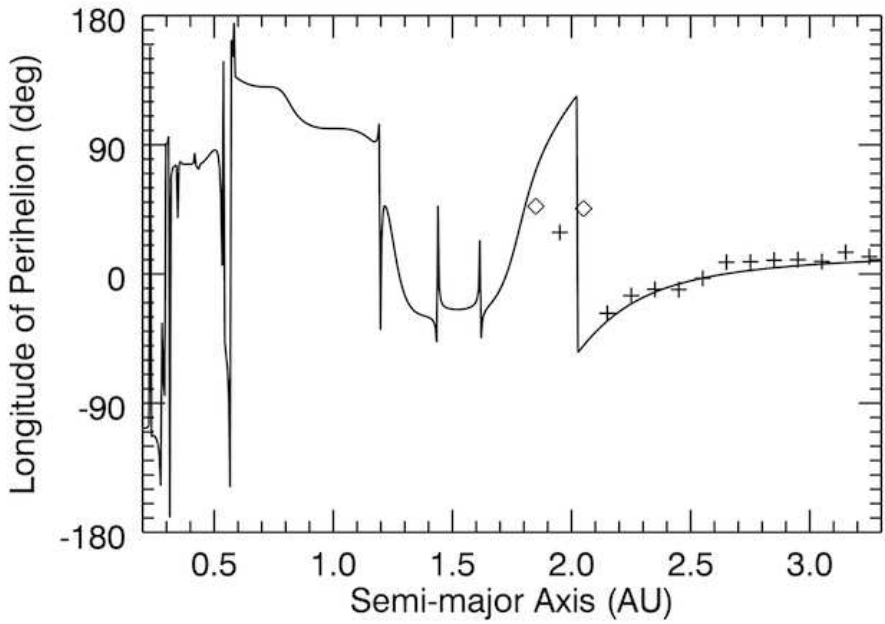}
  \caption{\sm The forced longitude of perihelion as a function of semimajor axis, $\varpi_{f} (a)$, as obtained from linear secular theory with eight planets, Mercury---Neptune.  The plus symbols indicate the statistically significant peak directions of $\varpi$ within each 0.1 AU semimajor axis bin, calculated for bright minor bodies having $H<15$ (see Table \ref{TB_LNP}). For those bins of low confidence level, peak directions are also calculated with a larger sample of a fainter magnitude cut-off, $H<19$; only two additional bins of statistical significance ($1-p<0.95$) are found, indicated with the diamond symbols.}
\label{FrcdLNP}
\end{figure}

\begin{table}
\centering
\caption{ Mean directions of minor bodies' $\overline\varpi (a)$ and their $p$--values, based on the Rayleigh z-test. All the data for minor bodies with limiting magnitude of $H<15$ is binned in 0.1~AU semimajor axis bins, for the range 1.7--3.3~AU. For two bins with $p>0.05$, peak directions were also calculated for the samples with $H<19$ and tabulated here. For $a\leq1.7$~AU, the sample sizes are small and no statistically significant peak directions are found.}
\begin{tabular}{| l | p{2cm} | l | l | l |}
\hline
  Bin &  absolute magnitude cut-off& Number & $\overline\varpi$ (deg) & $p$\\
  \hline  \hline
  $1.7<a<1.8$~AU & 15 & 10 & -17.1 & 7.40$\times 10^{-1}$ \\
  $1.8<a<1.9$~AU & 15 & 86 &  55.8 & 9.99$\times 10^{-2}$\\
                                  & 19 & 2484 & 47.2 & $< 10^{-3}$ \\ 
  $1.9<a<2.0$~AU & 15 & 273 & 28.9 & $< 10^{-3}$\\
  $2.0<a<2.1$~AU & 15 & 9 & 51.7 & 4.61$\times 10^{-1}$\\
                                  & 19 & 289 & 45.6 & 4.94$\times 10^{-3}$ \\ 
  $2.1<a<2.2$~AU & 15 & 787 & -27.5 & $< 10^{-3}$\\
  $2.2<a<2.3$~AU & 15 & 4147 & -15.1 & $< 10^{-3}$\\
  $2.3<a<2.4$~AU & 15 & 5593 & -10.8 & $< 10^{-3}$\\
  $2.4<a<2.5$~AU & 15 & 3805 & -10.9 & $< 10^{-3}$\\
  $2.5<a<2.6$~AU & 15 & 7113 & -3.1 & $< 10^{-3}$\\
  $2.6<a<2.7$~AU & 15 & 9370 & 8.0 & $< 10^{-3}$\\
  $2.7<a<2.8$~AU & 15 & 7618 & 8.4 & $< 10^{-3}$\\
  $2.8<a<2.9$~AU & 15 & 3029 & 9.4 & $< 10^{-3}$\\
  $2.9<a<3.0$~AU & 15 & 5618 & 9.8 & $< 10^{-3}$\\
  $3.0<a<3.1$~AU & 15 & 10681 & 8.6 & $< 10^{-3}$\\
  $3.1<a<3.2$~AU & 15 & 15016 & 15.0 & $< 10^{-3}$\\
  $3.2<a<3.3$~AU & 15 & 3772 & 12.0 & $< 10^{-3}$\\
\hline 
\end{tabular}
\label{TB_LNP}
\end{table}

The distribution of the longitude of perihelion, $\varpi$, of all detected NEOs was recently examined by \citet{Emelyanenko:2011}. These authors noted that the NEOs exhibit a preference of $\varpi$ near Jupiter's perihelion longitude, $\varpi_J \simeq 15^\circ$. We will show in this section that this preference is contributed only by the Amors, and that it is indeed owed to the effects of planetary perturbations. We will also show that the Atens and Apollos have distinctly different $\varpi$ distributions whose patterns are dominated by observational bias. 

\subsection{Amors}\label{ss:varpi-amors}
In this section, we first discuss the $\varpi$ distribution of all the known Amors, and then we examine the distributions of the bright and faint subsets.

The $\varpi$ distribution of all the Amors is shown as the red solid line in Figure~\ref{DistLNP}.  We observe a clear enhancement around the location of Jupiter's longitude of perihelion, $\varpi_J \simeq 15^\circ$: approximately 50\% more Amors have $\varpi$ within $\pm90^\circ$ of $\varpi_J$ than beyond that range.  (The $\chi^2$ value of this distribution relative to a uniform distribution is 76.4;  this exceeds $\chi^2_{crit}=35.2$ required to reject the uniform distribution.  The Rayleigh z-test also finds $p\ll10^{-3}$, confirming the statistical significance of this non-uniform distribution.)  A similar enhancement exists for the $\varpi$ distribution of the main belt asteroids, and was noted many decades ago and attributed to secular planetary perturbations~\citep{Kresak:1967,Tancredi:1998}.  
About 47$\%$ of Amors have semimajor axes in the range $2.1~\AU<a<3.3~\AU$; this is also the semimajor axis range where about 98$\%$ of main belt asteroids are located~(cf.~Figure~\ref{AstDistAE}).  Therefore, for comparison with the Amors, we plot as a black line in Figure~\ref{DistLNP} the $\varpi$ distribution of all main belt asteroids.  The similarity of the $\varpi$ distribution of the Amors (red line) to that of the MBAs is evident.  
In this figure, we also plot the $\varpi$ distribution of inner Amors located where the main belt asteroids are scarce (i.e., $a<2.1~\AU$); we observe that this distribution shows much less preference for values near $\varpi_J$.

Focussing on the bright Amors ($H<15$), we plot their $\varpi$ distribution as the red line in Figure~\ref{DistLNP2}).  We find that their mean longitude of perihelion is $\bar\varpi=4^\circ$.  The Rayleigh $z$--test indicates high statistical significance of this directionality, with a $p$--value of $\ll10^{-3}$.  Sixty-one percent of these objects have $\varpi$ values within $\pm 90^\circ$ of this peak direction, a $6.9\sigma$ departure from a uniform random distribution (adopting binomial statistics).  We further calculated the bright minor bodies' mean values of $\varpi$ in each 0.1 AU--wide bin in semimajor axis over the range 0.5--3.3~AU, and used the Rayleigh z-test to calculate their statistical significance.  For $a\leq1.7$~AU, the sample sizes in the 0.1~AU bins are very small, and we find no statistically significant directionality of $\varpi$. But we do find highly significant results for most bins at larger semimajor axes. These results are tabulated in Table~\ref{TB_LNP} and plotted in Figure~\ref{FrcdLNP}.  
We can trace the observed pattern (of mean $\varpi$ as a function of semimajor axis) to the effects of secular planetary perturbations which, over secular timescales, cause $\varpi$ to spend more time near the forced longitude of perihelion, $\varpi_f$, during an apsidal precession cycle.  This is illustrated as follows.  Using linear secular theory, as in \citet{Murray:1999}, we calculated the forced eccentricity vector $(e_f\cos\varpi,e_f\sin\varpi)$ for test particles; we included the secular perturbations of the eight major planets, Mercury through Neptune.  In Figure~\ref{FrcdLNP}, we plot the value of the forced longitude of perihelion, $\varpi_f$, as a function of semimajor axis.  We observe that, in the outer main belt ($2.5<a<3.3$~AU), $\varpi_f$ is smoothly and very slowly varying in the range $-5^\circ$ to $9^\circ$  but it drops rapidly to $-26^\circ$ and $-40^\circ$ as the semimajor axis decreases to 2.2 and 2.1 AU.  Comparing with the data, we find that the bright minor bodies' mean $\varpi$ roughly follows $\varpi_f$ values with semimajor axis.  We also confirmed that the mean $\varpi$ abruptly changes at $a\simeq2.0$~AU, as the secular perturbation theory predicts. Overall, secular perturbation theory explains the concentration of $\varpi$ near $\varpi_J$ for MBAs and Amors and also the weaker directionality of the $\bar\varpi (a)$ distribution of the inner Amors in Figure~\ref{DistLNP}.

In order to understand how observational bias contributes to the non-uniform $\varpi$ distribution, we examine the $\varpi$ distributions of the bright and faint Amors separately, and also the bright main belt asteroids.  These distributions are shown in Figure~\ref{DistLNP2}.  We observe that the dominant feature in all the curves is the peak near $\varpi\approx\varpi_J$. In addition to the $\varpi$ concentration due to the secular dynamics, there are several less dominant features that are likely owed to observational bias. The faint Amors' $\varpi$ distribution exhibits a deficit near $\varpi\approx-90^\circ$; this feature is not present in the bright Amors' $\varpi$ distributions.  The deficit of faint Amors at $\varpi\approx-90^\circ$ is related to the seasonal unfavorable observational conditions when the Galactic center and the southwestern US Summer monsoon season are near the opposition longitude, i.e., when $\lambda_{op}\approx-90^\circ$, as discussed in Section~\ref{s:nodes}.  This causes a deficit in discoveries of those faint asteroids whose perihelia are located close to this opposition longitude, i.e., near $\varpi \simeq -90^\circ$. The semimajor axis distribution of faint Amors also causes additional observational bias. This is because, in order to be observable, faint Amors must have smaller $a$ than the bright Amors.  This explanation is supported by the data: the faint Amors ($H > 25$) have median semimajor axis of 1.6~AU whereas the bright ones ($H < 19$) have median semimajor axis of 2.2~AU. The small $a$ population is more uniformly distributed as can be seen in Figure~\ref{DistLNP}, which hides the concentration feature at $\varpi \simeq \varpi_J$.

\subsection{Atens and Apollos}\label{ss:varpi-atens-apollos}

\begin{figure}
\centering
 \includegraphics[width=6in]{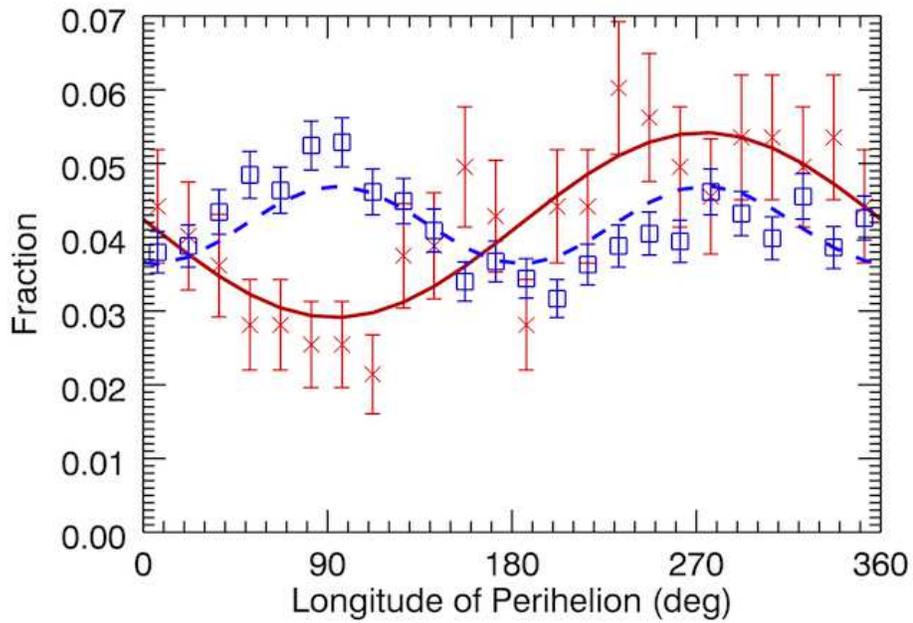}
 \caption{\sm Distribution of the longitude of perihelion, $\varpi$, of Atens (red points, N=747) and Apollos (blue points, N=4767). The smooth curves indicate the best-fit sinusoidal functions of $2 \pi$ period for Atens and $\pi$ period for Apollos. }
\label{LNPECA}
\end{figure}

\begin{figure}[h]
\centering
 \includegraphics[width=220px]{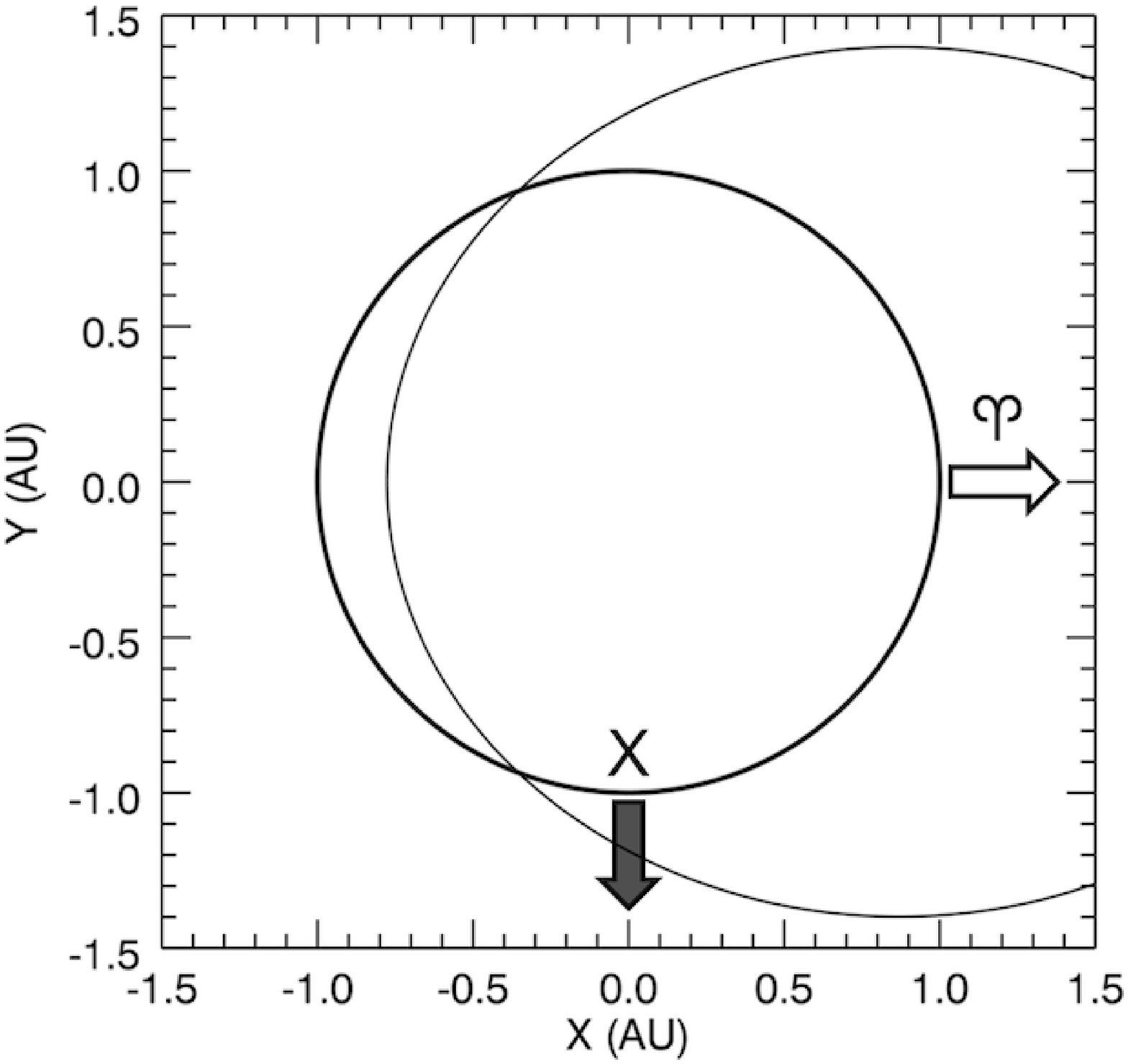}
 \includegraphics[width=220px]{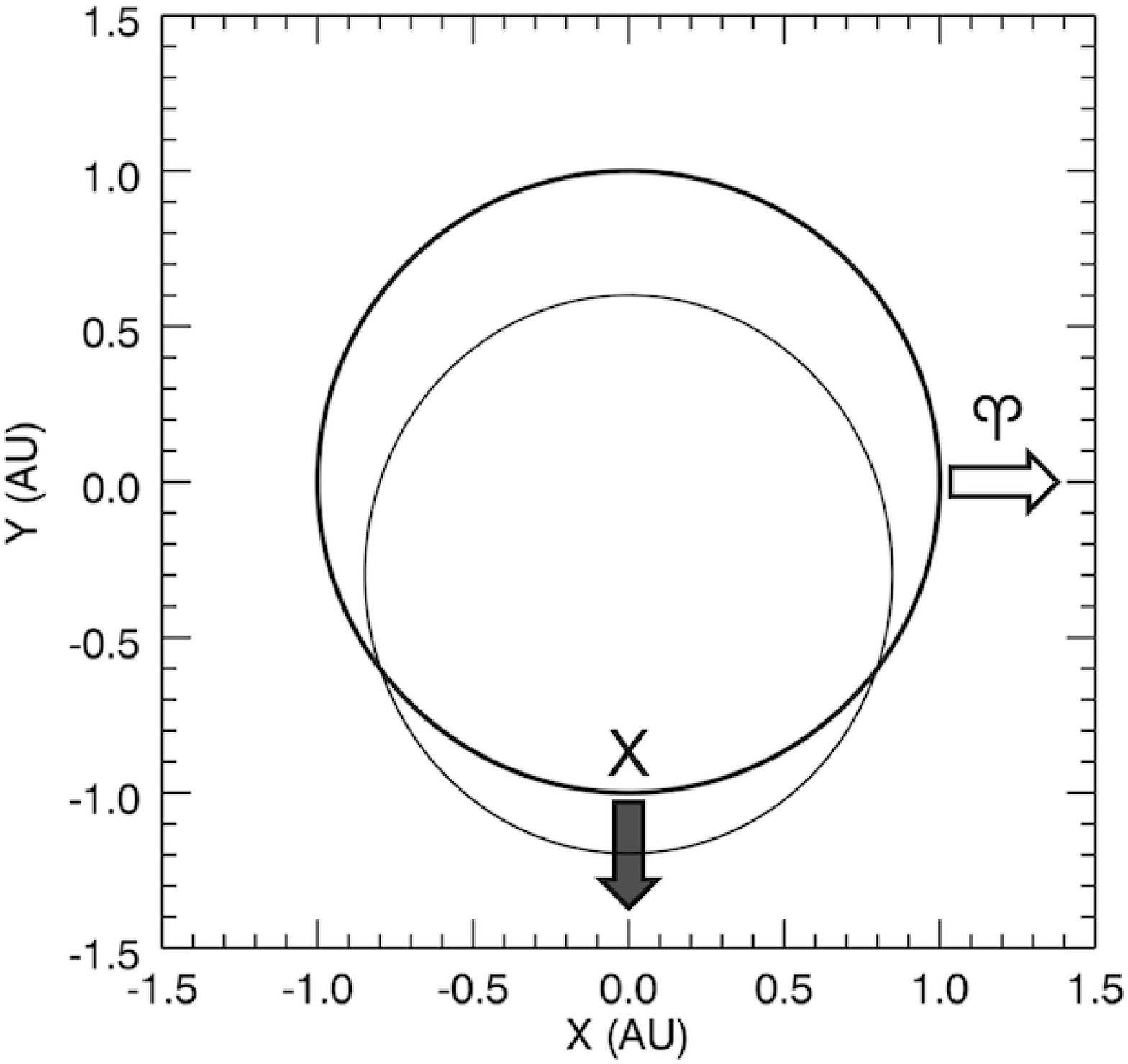}
\caption{\sm Schematic diagrams of observationally unfavorable configurations for Apollos (left panel) and Atens (right panel). The Sun is located at the center and the orbit of Earth is shown as the circle with 1~AU radius.  The zero longitude where Earth passes during September is located at the positive x-direction and indicated with the open arrows. The direction of Galactic center crossing is located in the negative y-direction and indicated with the symbol `X' and the solid black arrows. Note that the $180^\circ$--rotated configuration of the example for Apollos (left panel) is also observationally unfavorable.}
\label{SchD_ECA}
\end{figure}

The $\varpi$ distributions of the Earth--crossing subgroups of NEOs, the Atens and the Apollos, are significantly different from that of the Amors and also from each other.   First, we examined the 70 bright Atens and the 948 bright Apollos ($H<19$).  The Rayleigh z--test finds $p$--values well in excess of 0.05, indicating no directionality in these $\varpi$ distributions.   We also binned the data in 24 bins, each 15$^\circ$ wide, and calculated the $\chi^2$ values with respect to a uniform distribution; the bright Atens and bright Apollos have $\chi^2$ values of 12.3 and 13.7, respectively, indicating that these $\varpi$'s are consistent with uniform random. ($\chi^2 > 35.2$ is required to reject the uniform distribution.)  The lack of non-uniformity is remarkable considering the significant non-uniformity of the bright ($H<19)$ Amors' $\varpi$ distribution (see section 5.1). 

The $\varpi$ distributions of all the Atens and Apollos are plotted in Figure~\ref{LNPECA}.  In some contrast with the bright subsets, the $\varpi$ distributions of the full sets of Atens and  Apollos deviate significantly from a uniform distribution: we find $\chi^2$ of 49.4 and 80.0, and $p<0.005$ and $p<0.001$, respectively, relative to a uniform distribution. We can reject the uniform distribution with high confidence level for both the Atens and the Apollos.  The Atens concentrate around $\varpi=270^\circ$, whereas the Apollos exhibit double peaks at $90^\circ$ and $270^\circ$.  We also used the Rayleigh z--test and confirmed the unimodal distribution for Atens and the bimodal (axial) distribution of Apollos, each with high confidence level of $p\ll10^{-3}$. The Atens' peak direction is $274^\circ$, and 59\% of the Atens have $\varpi$ within the half-circle centered at this value.   The double peaks of the Apollos are at $95^\circ$ and $275^\circ$, and approximately 54\% of Apollos have $\omega$ within the two quadrants centered at these two peaks.

In Figure~\ref{LNPECA}, we also plot the best fit sinusoidal functions ($2\pi$ period for the Atens, and $\pi$ period for the Apollos).  (These fitted functions are shown merely to illustrate that the observational data show the $\pi$ and $2\pi$ periodicities; we do not claim that the distributions actually have simple sinusoidal functional forms.)  Since these non-uniform patterns are absent in the bright subsets, it is likely that they are owed to observational bias. Earth--crossing asteroids are brightest when they are near 1~AU heliocentric distance {\it and} at opposition, where they present a low phase angle to Earth--based observers.  When this direction coincides with observationally unfavorable locations, these objects are not easily detectable.  At these locations, the orbital geometry presented by the Apollos is different than for the Atens.  We illustrate this in the schematic diagrams in Figure~\ref{SchD_ECA}. In this figure, the observationally unfavorable regions are located along the positive y direction (Galactic crossing during northern Winter) and along the negative y direction (Galactic crossing during northern Summer).  For the orbital geometry of Apollos (left panel), this means that objects having $\varpi\approx0$ or $\varpi\approx180^\circ$ are difficult to observe.  (A similar approach was used by \citet{Valsecchi:1999} to explain the selection effects in the so-called ``Taurid" group of Apollo asteroids.) For the Atens (right panel), 
the detection of objects having $\varpi\approx90^\circ$ is disfavored because their sky position at opposition is aligned with the Galaxy.  These observational biases due to the specific orbital geometries of these NEO subgroups explain qualitatively the the most prominent features in the non-uniform distribution of  $\varpi$ of the Apollos and the Atens in Figure~\ref{LNPECA}. 

A less prominent but also statistically significant feature for the Apollos is that their peak near $\varpi\approx90^\circ$ is larger than the one near $\varpi\approx270^\circ$.  There are 2493 (52.3\%) Apollos having $\varpi$ in the half-circle centered at 90$^\circ$ compared to 2274 (47.7\%) in the complementary half-circle; this is a $3.2\sigma$ departure from a uniform random distribution (adopting binomial statistics).  Although this is reminiscent of a similar feature in the distribution of the longitude of ascending nodes of NEOs~(Section~\ref{s:nodes}), unlike the case for the nodes, the known observational selection effects do not preserve a $\pi$--periodicity for $\varpi$.  The observational conditions for faint objects are worse during Northern Summer compared to those in the Northern Winter (positive y direction and negative y direction, respectively, in Fig.~\ref{SchD_ECA}).  In addition, the Monsoon season of the southwestern United States, which occurs during the months of July--September, makes observational conditions worse over the fourth quadrant in Fig.~\ref{SchD_ECA}.  Combined with the orbital geometry of the Apollos, these effects qualitatively account for the slightly larger peak of $\varpi$ near 90$^\circ$ compared to the peak near $270^\circ$.  We leave the complicated quantitative modeling of these observational selection effects as a subject for a future study.

\section{Summary and Conclusions}\label{s:conclusions}
 
The distributions of the angular elements (the longitude of node, argument of perihelion and longitude of perihelion) of the known NEOs are strongly non-uniform.  We have considered the observational biases (due to seasonal effects and the relative geometry between Earth observers and NEOs) and the dynamical effects of secular planetary perturbations that may cause these non-uniformities.  Our main results are summarized as follows.

\begin{enumerate}

\item
The apparent distribution of the longitude of ascending node, $\Omega$, is strongly affected by observational biases.  However, the lack of $\pi$ periodicity in this distribution cannot be explained by known selection effects, and indicates an intrinsic non-uniformity with a concentration near $\Omega=111^\circ$, approximately coinciding with Jupiter's longitude of ascending node, $\Omega_J=100^\circ$.  
There are $\sim53\%$ NEOs in the half-circle centered at $111^\circ$, a 5.5$\sigma$ departure from a uniform random distribution. For comparison, we find that there are $\sim 56\%$ main belt asteroids with $\Omega$ values in the half-circle centered at $94.5^\circ$, an $82\sigma$ departure from a uniform random distribution.  Previous studies have attributed this preference to planetary secular perturbations and the approximate coincidence of the mean plane of the main belt asteroids with Jupiter's orbital plane or the solar system's invariable plane.  Our result for the NEOs implies that secular planetary perturbations cause the NEOs' mean plane to deviate from the ecliptic.  Direct evidence from the observationally nearly complete sample of bright NEOs ($H<19$) is statistically marginal, although their mean value, $\bar\Omega=115^\circ$, is also similarly close to Jupiter's.  We predict that the intrinsic non-uniform $\Omega$ distribution of the NEOs will be directly revealed in the near future when the observationally complete sample size exceeds $\sim2500$.

\item 
The three subgroups of NEOs (the Amors, the Atens and the Apollos) have distinctly different and non-uniform apparent distributions of the argument of perihelion, $\omega$.  We find that the different Earth--NEO geometries presented by the three subgroups account for some of these differences.  For the Amors, the intrinsic distribution of $\omega$ appears to be consistent with a nearly uniform random distribution; any intrinsic non-uniformity is below the statistical errors in the data.  However, for the intrinsic $\omega$ distribution of the Apollos, we find a statistically significant deviation from a uniform random distribution of $\omega$. The distribution is axial, with enhancements near $\omega\simeq 0$ and $180^\circ$.  Approximately 55\% of bright Apollos have $\omega$ within the two quadrants centered at 0 and $180^\circ$, a $3.2\sigma$ departure from a random distribution.  We attribute this non-uniformity to the Kozai effect arising from Jovian perturbations.  It is intriguing that the Amors do not show evidence of this dynamical effect but the Apollos do; we leave this question for a future study.

\item 
The Amors exhibit significant non-uniformity in their apparent distribution of the longitude of perihelion, $\varpi$.  We find that their intrinsic $\varpi$ distribution (based on the bright objects, $H<19$) is also non-uniform with high statistical significance; they peak near $4$~degrees ecliptic longitude, close to Jupiter's longitude of perihelion.  Sixty-one percent of bright Amors have $\varpi$ values in the half-circle centered at 4~degrees; this is a $6.9\sigma$ departure from a uniform random distribution.  We find that the directionality of the intrinsic $\varpi$ distribution of all asteroids, including Amors, varies with semimajor axis, and is consistent with secular planetary perturbations.

\item The Atens exhibit an apparent concentration near $\varpi=274^\circ$, with approximately 59\% Atens having $\varpi$ in the half-circle centered at this value.  The Apollos' apparent distribution of $\varpi$ is axial, with two peaks centered at $\varpi=+95^\circ$ and $\varpi=275^\circ$, with approximately 54\% Apollos having $\varpi$ values in the two quadrants centered at these peak values.  We find that these different patterns are owed to observational selection effects due to the different geometries presented by these subgroups.  The intrinsic distribution of $\varpi$ for the Atens and Apollos (based on the sample of bright objects, $H<19$) does not show the non-uniform pattern seen in the Amors, and is  consistent with a uniform random distribution.

\end{enumerate}

Our results show that, despite their strongly chaotic dynamics, the NEOs' angular elements, $\Omega$ and $\omega$ and $\varpi$, have a modest but statistically significant level of non-uniformity due to planetary perturbations. Recent models of the NEOs' orbital distribution \citep{Bottke:2002,Greenstreet:2012}, have reported on only three orbital parameters, semimajor axis, eccentricity and inclination.  We suggest that the angular elements, $\Omega$, $\omega$, and $\varpi$ should be added to this orbital element set for more accurate representation of the distribution of NEOs.

\acknowledgements

We thank Takashi Ito for discussions and for providing simulation data. This research was supported by NSF grant \#AST-1312498.

\bibliographystyle{icarus}

\bibliography{mybib}

\end{document}